# Quantum spin driven Yu-Shiba-Rusinov multiplets and fermion-parity-preserving phase transition in K$_3$C$_{60}$

Shu-Ze Wang[1†], Xue-Qing Yu[1†], Li-Xuan Wei[1], Li Wang[2], Qiang-Jun Cheng[1], Kun Peng[2], Fang-Jun Cheng[1], Yu Liu[1], Fang-Sen Li[2*], Xu-Cun Ma[1,3], Qi-Kun Xue[1,3,4,5*], Can-Li Song[1,3*]

[1]*State Key Laboratory of Low-Dimensional Quantum Physics, Department of Physics, Tsinghua University, Beijing 100084, China*

[2]*Vacuum Interconnected Nanotech Workstation, Suzhou Institute of Nano-Tech and Nano-Bionics, Chinese Academy of Sciences, Suzhou 215123, China*

[3]*Frontier Science Center for Quantum Information, Beijing 100084, China*

[4]*Beijing Academy of Quantum Information Sciences, Beijing 100193, China*

[5]*Southern University of Science and Technology, Shenzhen 518055, China*

Magnetic impurities in superconductors are of increasing interest due to emergent Yu-Shiba-Rusinov (YSR) states and Majorana zero modes for fault-tolerant quantum computation. However, a direct relationship between the YSR multiple states and magnetic anisotropy splitting of quantum impurity spins remains poorly characterized. By using scanning tunneling microscopy, we resolve systematically individual transition-metal (Fe, Cr and Ni) impurities induced YSR multiplets as well as their Zeeman effects in K$_3$C$_{60}$ superconductor. The YSR multiplets show identical *d* orbital-like wave functions that are symmetry-mismatched to the threefold K$_3$C$_{60}$(111) host surface, breaking point-group symmetries of the spatial distribution of YSR bound states in real space. Remarkably, we identify an unprecedented fermion-parity-preserving quantum phase transition between ground states with opposite signs of the uniaxial magnetic anisotropy that can be manipulated by an external magnetic field. These findings can be readily understood in terms of anisotropy splitting of quantum impurity spins, and thus elucidate the intricate interplay between the magnetic anisotropy and YSR multiplets.

†These authors contributed equally to this work.

*Correspondence to: clsong07@mail.tsinghua.edu.cn, fsli2015@sinano.ac.cn, qkxue@mail.tsinghua.edu.cn

**1. Introduction**

Magnetic adsorbates on solid surfaces hold promise for creating and manipulating quantum many-body phenomena such as Kondo singlets [1] and Yu-Shiba-Rusinov (YSR) states [2]. Recently, there has been a revival of interest in the YSR states, partially fueled by a viable proposal of YSR chains for realization of Majorana-based quantum computation [3,4]. In classical spin-$S$ and quantum spin-1/2 models [5-7], the YSR excitations from a many-body ground state to excited state form a single pair of bound states inside the superconducting gaps, yet multiple YSR pairs have been observed on transition-metal impurities [8-13]. Such a dichotomy arouses renewed attention on quantum impurity models for higher spins $S$ [14,15], which predict multiple quantum phase transitions (QPT) including the previously explored one between free-spin and Kondo-screened ground states with different fermion parity, controlled by an exchange coupling $J$ of the impurity spin $S$ with the superconductors [16]. Despite this advance, the YSR multiplets have been diversely assigned to scattering channels with different angular momenta [8,17] or orbitals [9-11], or to magnetic anisotropy – a property that confers a preferential direction on the impurity spins [12,13].

For transition-metal adatoms, the higher quantum spins up to 5/2 are subject to unique SU(2) symmetry breaking by uniaxial ($D$) and transverse ($E$) magnetic anisotropies [18-22]. As the Zeeman splitting by an external magnetic field $B$ is included, the spin Hamiltonian can be written as

$$H_{\text{eff}} = DS_z^2 + E(S_x^2 - S_y^2) + g\mu_B \boldsymbol{B} \cdot \boldsymbol{S}, \qquad (1)$$

where $\boldsymbol{S} = (S_x, S_y, S_z)$ is the spin operator, $g$ and $\mu_B$ represent the Landé factor and Bohr magneton, respectively. In this situation, the number of YSR bound states becomes sensitive to the ground state of quantum spins and the magnetic anisotropy [13-16]. However, their interplay remains to be elucidated in experiment. To observe the spin-related fine structure of the YSR bound states, the magnetic anisotropy should be small so that the superconducting gap $\Delta$ can accommodate the resulting YSR multiplets. Provided the typical $\Delta \sim 1.3$ meV in usual Pb, Nb and NbSe$_2$ host superconductors [2], the experiment is challenging because very high energy resolution of several tens $\mu$eV is required to resolve the tiny anisotropy splitting [23]. Recently, two studies of manganese phthalocyanine molecules ($S = 1$) on Pb and Fe vacancies ($S = 2$) on (Li$_{0.8}$Fe$_{0.2}$)OHFeSe reported signatures of the magnetic anisotropy resulted YSR multiplets [12,13]. However, the key experimental evidences of the quantum spin driven YSR multiplets – identical spatial wave function for various YSR states and fermion-parity-preserving QPT [15], are wholly missing at present. Here we report systematic measurements of the transition-metal adatoms with different spins, namely Fe$^{3+}$ ($S = 5/2$), Cr$^{2+}$ ($S$

= 2) and $Ni^{2+}$ ($S = 1$), on superconducting $K_3C_{60}$(111) by means of a cryogenic scanning tunneling microscopy (STM) (Fig. 1a). We directly visualize the long-sought $d$ orbital-derived identical YSR wave functions for various YSR multiplets on Fe adatoms that change little with Zeeman splitting. These results unambiguously established the quantum nature of high-$S$ impurities, which enables us to identify an unprecedented fermion-parity-preserving QPT on Ni. Such a QPT connects two distinct spin ground states with opposite signs in $D$, which can be straightforwardly tuned by an external magnetic field.

## 2. Materials and methods

Our experiments were conducted in two ultrahigh vacuum (UHV) cryogenic (down to 0.4 K) scanning tunneling microscopy systems, integrated with respective molecular beam epitaxy (MBE) chambers for *in-situ* sample preparation and superconducting magnets perpendicular to the sample surface up to 9 T and 11 T. The base pressure of all STM and MBE chambers is lower than $2.0 \times 10^{-10}$ Torr. High-purity $C_{60}$ molecules were evaporated from standard Knudsen cells and grew layer-by-layer on nitrogen-doped SiC(0001) wafers (0.1 Ω·cm) at 200°C, which were pre-graphitized by thermal heating to form the bilayer graphene-dominant surface. Potassium (K) atoms were then deposited on the $C_{60}$ epitaxial films at a low temperature of ~ 200 K step by step, followed by post-growth annealing at room temperature. We fixed the $C_{60}$ film thickness to five monolayers and the K doping close to three in this study.

In order to systematically investigate individual impurities in fulleride superconductors, three magnetic (Fe, Cr and Ni) atoms have been evaporated from their respective Knudsen cells on the $K_3C_{60}$ samples at low temperature (~ 150 K). This effectively reduces the diffusion and leads to formation of individual impurity atoms. All spectroscopic measurements were conducted by disenabling the feedback circuit, sweeping the sample bias voltage $V$, and recording the differential conductance using a standard lock-in technique with a small bias modulation (0.1 mV) at 983 Hz. Polycrystalline PtIr tips were cleaned by *e*-beam bombardment in UHV chamber and appropriately calibrated on Ag/Si(111) epitaxial films prior to each STM measurement. The STM topographies were measured in a constant current mode.

## 3. Results and discussion

The *s*-wave superconductor $K_3C_{60}$ has a maximum gap of 5.4 meV [24], albeit spatial inhomogeneity. The large gap ensures the quantum spin driven YSR multiplets accessible. Additionally, its high upper critical field of 90 telsa enables us to detect the YSR states under a sizable Zeeman splitting when the system remains in the superconducting state [25]. We first deposited Fe on $K_3C_{60}$(111) that individually occupy the top (Fe(I)) and near-hollow (Fe(II)) sites of the hexagonal lattice at the top $C_{60}$ molecules (Fig. 1b and c). Owing to site-specific interactions with $K_3C_{60}$, the differential conductance spectra d$I$/d$V$, proportional to the local density of states (DOS) of the sample surface, on Fe(I) and Fe(II) are characterized by distinctly different YSR states. Two (marked by α ~ 0.18 meV and β ~ 0.48 meV) and three (marked by δ ~ 0.64 meV, γ ~ 1.19 meV and ε ~ 1.97 meV) pairs of YSR states are revealed on Fe(I) (Fig. 1d) and Fe(II) (Fig. 1e), respectively. The site-sensitive YSR states are also evidenced by modifying the registry of the same Fe(II) in Fig. 1e, a tiny variation of which makes a remarkable variation in the YSR energy $E_{YSR}$ (Supplementary Fig. 1). Note that the energy spacing between the YSR states ±α is so small that they coalesce into a prominent zero-bias conductance peak at the center of Fe(I) (Supplementary Fig. 2). This indicates that $J$ is close to the parity-symmetry-breaking QPT point for the top-site Fe(I) adatoms [2,16,26].

**Table 1. YSR energies on Fe(I) and Fe(II) adatoms without and with an external magnetic field $B$ = 7 T.** The values in brackets denote the $E_{YSR}$ of the Fe(II) impurity atom after the thermal process. The errors correspond to the standard deviations from the Gaussian fits.

| $E_{YSR}$ (meV) | α | β | δ | γ | ε |
|---|---|---|---|---|---|
| 0 T | 0.187 ± 0.009 | 0.483 ± 0.034 | 0.640 ± 0.059 (1.926 ± 0.021) | 1.192 ± 0.002 (0.822 ± 0.002) | 1.986 ± 0.011 (2.436 ± 0.069) |
| 7 T | 0.007 ± 0.004 | 0.941 ± 0.013 | 0.917 ± 0.016 (2.240 ± 0.024) | 1.458 ± 0.010 (1.027 ± 0.001) | 2.091 ± 0.057 (/) |
| Shift | 0.180 ± 0.013 | 0.458 ± 0.044 | 0.277 ± 0.075 (0.314 ± 0.045) | 0.266 ± 0.012 (0.205 ± 0.003) | 0.105 ± 0.068 (/) |

By applying a magnetic field of 7 T normal to the sample surface, we reveal a few hundreds $\mu$eV energy shifts of the YSR states on both Fe(I) and Fe(II) (marked by the black arrows in Fig. 1d and e and Supplementary Fig. 1b). Gaussian fits of the YSR peaks (red curves) allow us to extract every $E_{YSR}$ and field-induced shifts, as summarized in Table 1. We here ascribe the observed YSR multiplets to magnetic anisotropy splittings of the higher-spin quantum impurity. This claim is supported by our simultaneous observation of inelastic spin-flip excitations [18-22], which develop as discrete conductance peaks outside the superconducting gaps (Supplementary Fig. 3). A quantitative analysis reveals two spin-flip excitations at energies of ~ 1.8 meV and ~ 3.6 meV that have a twofold difference, based on which we deduce the $S$ = 5/2 spin state for Fe adatoms [19]. Only the spin-5/2 multiplet splits under a magnetic anisotropy $D$ into three doublets, |5/2, ±1/2⟩, |5/2, ±3/2⟩ and |5/2, ±5/2⟩, separated by the observed excitation energies of 2|$D$| and 4|$D$| (Supplementary Fig. 3c). This shows that every Fe loses three electrons (two from the 4$s^2$ and one from

the $3d^6$) and an oxidation state of $Fe^{3+}$ is formed, thanks to the high electron affinity of $C_{60}$. As thus, the two pairs of YSR states can be readily understood in terms of a partially screened ground state and an easy-axis anisotropy $D < 0$ on Fe(I) (Fig. 1f, Supplementary Section 1, Supplementary Figs. 4 and 5). At 0 T, the spin selection rule ($\Delta S_z = \pm 1/2$) allows two excitations from the doublet ground state $|2, \pm 2\rangle$ to excited states $|5/2, \pm 5/2\rangle$ and $|5/2, \pm 3/2\rangle$, giving the two YSR states α and β, respectively. The field $B$ splits spin states with same $|S_z|$ by the Zeeman term $g\mu_B B S_z$ and shifts the YSR pair ±α (±β) towards (away from) $E_F$ by the $S_z$-dependent Zeeman splitting.

In contrast, three pairs of YSR states must imply a non-zero transverse anisotropy $E$ that mixes states of different $S_z$ on Fe(II) (for details see Supplementary Section 1). This matches with its asymmetric adsorption site on $K_3C_{60}(111)$. Since all visible YSR excitation states (δ, γ and ε) merely shift away from $E_F$ at 7 T (Fig. 1e and Supplementary Fig. 1b), the Fe(II) instead denotes a free-spin ground state ($S = 5/2$) with $D < 0$, which favors a singlet ground state $|\Psi_{5/2}^0\rangle$ with predominant weights at large $|S_z|$ values (Fig. 1g). Otherwise, the magnetic field would shift some YSR states toward $E_F$. In principle, five pairs of YSR states arise from $|\Psi_{5/2}^0\rangle$ to $|\Psi_2^n\rangle$ ($n = 0, 1, 2, 3, 4$), yet two of them might merge into the continuum of quasiparticle states due to the large $E_{YSR} > \Delta$ (Fig. 1g).

Importantly, our spectroscopic $dI/dV$ maps show identical spatial wave functions for the YSR multiplets, irrespective of the Zeeman splitting. This result renders a conclusive confirmation of the quantum spin driven YSR multiplets. At 4.2 K, the YSR maps are characteristic of pseudo-fourfold symmetric quasiparticle clouds that look as clover-leaf shaped $d$ orbitals with four lobes (Fig. 2a, Supplementary Figs. 6 and 7,). The deviation from perfect $d$ orbitals is phenomenally resulted from the apparent symmetry mismatch between the $d$ orbitals and threefold chemical environment of $K_3C_{60}(111)$. One nodal plane of $d$ orbital-like YSR maps ($d_{x^2-y^2}$ or $d_{xy}$) runs along the close-packed directions of $C_{60}$, while adjacent quasiparticle lobes coalesce along the other nodal plane. The resultant breaking of the fourfold rotational symmetry becomes so pronounced that dimer-shaped (Fig. 2b and c, Supplementary Figs. 8 and 9) YSR patterns are discernible at 0.4 K. Note that the dimer-shaped YSR pattern appears to be more localized than the $d$ orbital-like YSR maps at 4.7 K, which primarily correlates with the sharply increased spatial dependence of the YSR intensity at 0.4 K. Anyhow, the YSR maps display essentially identical spatial distribution for various YSR excitations on the respective Fe(I) and Fe(II) adatoms, a pivotal hallmark of quantum spin driven YSR multiplets. Note that the asymmetric intensities of the particle- and hole-like YSR states are originated from a local potential scattering $U$ by the impurities [8-13,18,27].

To our best knowledge, the unique $d$ orbital symmetry fingerprints of individual magnetic adatoms are imaged for the first time on the $C_{3v}$-symmetric surface with $s$-wave superconductivity. This finding virtually eliminates alternative scenario of either $d$-wave superconducting gap [28,29] or anisotropy of the projected Fermi surface [2,9,30], which have been suggested as origins of the anisotropic YSR patterns. In Fig. 2d and e, we plot two representative series of d$I$/d$V$ spectra across the Fe(I) and Fe(II) measured at 0.4 K, respectively. Interestingly, the YSR states ±α are detached away from the Fe(I) impurity site and all $E_{YSR}$ oscillate in real space. At 7 T, the ±α pair shifts towards $E_F$ and recoalesces into robust zero-energy peaks (Supplementary Fig. 10), whereas other YSR pairs (β, δ, γ and ε) shift in the opposite direction (Supplementary Figs. 10 and 11). Gaussian fits of the YSR peaks allow to quantify their spatial evolutions, which unexpectedly break the radial symmetry with respect to the impurity sites (Supplementary Fig. 12).

The spatially oscillating $E_{YSR}$ cannot be accounted for by STM tip-induced variations of the electrostatic potential or exchange coupling $J$ [20,31,32], consistent with the invariance of the $E_{YSR}$ with the tunnelling distance (Supplementary Fig. 13). Considering that $E_{YSR}$ relies on the normal-state electronic density of states at $E_F$ [2], a straightforward explanation of the spatial oscillation of $E_{YSR}$ involves Friedel-like screening off impurities, which modulates the local DOS $\rho(r)$ at $E_F$ and consequently $E_{YSR}$. Contrasting with point scatterer models [2,33], however, the crystal field splitting of $d$ orbitals results in orbital-selective scattering potential and exchange coupling $J$ with $K_3C_{60}$ [9-11]. Therefore, individual Fe impurities should be best regarded as extended scatterers with $d$-orbital structures. This anisotropy is mismatched with the $C_{3v}$ symmetry of the $K_3C_{60}$(111) surface, which breaks all point-group symmetries of the YSR states, as observed.

It is worth stressing that the $E_{YSR}$ on each Fe adatom displays synchronous variations against space and magnetic field (Supplementary Fig. 12). This supports the magnetic anisotropy origin of the YSR multiplets. One exception is the preferential coalescence of the YSR pair ±α to zero energy in the magnetic field. Note that the ±β shift of 0.45 ± 0.07 meV is consistent with the Zeeman splitting difference (0.41 meV) between the ground state $|2, ±2\rangle$ and excited state $|5/2, ±3/2\rangle$ at 7 T (Table 1). Intuitively, one expects the same amount of energy shift for ±α, in contrast to our experiment (Supplementary Fig. 12a). This zero-bias pinning should correlate with the quantum many-body effects [34], which merits further investigations.

Next, we study another transition-metal Cr adatoms, which occupy either hollow sites (Fig. 3a) or bridge sites (Fig. 3b) on the $K_3C_{60}$(111) surface. Although both adatoms cause spin-flip excitations, only the hollow-site Cr induces one YSR pair near ± 2.05 meV labeled as ±η (Fig. 3c and d). Distinct from Fe, the two spin-flip excitations at ~ 1.2 meV and ~ 3.6 meV have a threefold difference in energy, suggesting a $S = 2$ spin state of the $Cr^{2+}$ adatoms in a $d^4$ electron configuration (Fig. 3e). At 7 T, each YSR peak is split in two separated by the Zeeman energy of 0.81 meV (Fig. 3f). The findings uncover an easy-plane magnetic anisotropy $D$ ~ 1.25 meV for Cr and the YSR states ±η arise from excitations from a singlet ground state |2, 0⟩ to doublet excited state |3/2, ±1/2⟩, as sketched in Fig. 3g. On Cr, we note that the spatial variation of ~ 0.05 meV in $E_{YSR}$ (Supplementary Fig. 14a) is smaller than those of Fe(I) (0.23 meV) and Fe(II) (0.15 meV). Nevertheless, the spatial evolutions of $E_{YSR}$, the intensity and particle-hole asymmetry are all no longer radially symmetric (Supplementary Fig. 14).

To provide a more complete picture of the magnetic anisotropy effects on the YSR states, we investigate the $S = 1$ spin state of $Ni^{2+}$ in its $d^8$ electron configuration. This renders it easy to conduct quantitative analyses of the Zeeman effects on $E_{YSR}$ even with non-zero transverse anisotropy $E$ (Supplementary Section 1), which does occur for randomly distributed Ni adatoms on $K_3C_{60}$ (Supplementary Fig. 15). Nevertheless, they favor a free-spin ground state (Supplementary Fig. 16 and Section 2) and consistently yield one pair of YSR peaks just as Cr at 0 T, but showing a huge variation in $E_{YSR}$ (Supplementary Fig. 17). Diagonalization of the $S = 1$ Hamiltonian in Eq.1 gives the eigenstates $|\Psi_1^0\rangle$, $|\Psi_1^-\rangle$ and $|\Psi_1^+\rangle$, with the latter two denoting the mixed states from |1, ±1⟩ by $E$ (Supplementary Section 1). Apparently, the free-spin ground state, either $|\Psi_1^0\rangle$ or $|\Psi_1^-\rangle$, relies on the sign of the $D$ and its relative magnitude to $\sqrt{E^2+(g\mu_B B)^2}E$. The particle-hole asymmetry in the YSR intensity can be reversed between different Ni adatoms (Supplementary Fig. 17). As the screened ground state for $S = 1$ generally creates two pairs of YSR peaks at 0 T (Supplementary Fig. 5), this finding implies that there must exist a sign change in $U$ for various Ni impurities with opposite particle-hole asymmetry of the YSR intensity (Supplementary Section 2) [27].

Irrespective of the sign change in $U$, the diversity of Ni adsorption sites allows to identify three distinct Zeeman effects on $E_{YSR}$, as illustrated in Fig. 4 and Supplementary Fig. 18. At $D > 0$, $|\Psi_1^0\rangle$ lies in the ground state and each YSR peak is split into two due to a lifting of the Kramers' doublets of the excited |1/2, ±1/2⟩ state by $B$ (Fig. 4a and Supplementary Fig. 18a). A linear fit of the $B$-dependent $E_{YSR}$ shown in Fig. 4b and

Supplementary Fig. 18b yields $g = 2.04 \pm 0.04$. By contrast, the YSR peaks are observed to mainly shift away from $E_F$ for $D < 0$ and smaller $E$, where $|\Psi_1^-\rangle$ is the ground state (Fig. 4c). The shift derives from the Zeeman splitting difference between $|\Psi_1^-\rangle$ and the excited state $|1/2, \pm1/2\rangle$, which reads as

$$\Delta E_{\text{YSR}} = \sqrt{E^2 + \left(g\mu_B B\right)^2} - E \pm \frac{1}{2}g\mu_B B. \qquad (2)$$

In principle, the excitation from $|\Psi_1^-\rangle$ to the $|1/2, 1/2\rangle$ state is allowed as well. However, the magnetic field $B$ considerably suppresses the weight of $|1, 1\rangle$ in $|\Psi_1^-\rangle$ for smaller $E$ (see inset in Fig. 4d) and renders only the other YSR excitation from $|1, 1\rangle$ to $|1/2, -1/2\rangle$ discernible in Fig. 4c. Using $g = 2$, the best fit of $\Delta E_{\text{YSR}}$ to Eq. 2 indeed yields a tiny $E = 0.09 \pm 0.01$ meV in Fig. 4d.

Most remarkably, we observe significant discontinuities in $E_{\text{YSR}}$ as a function of $B$ on some Ni impurities (Fig. 4e, Supplementary Fig. 18c and e). They turned out to be originated from a QPT between the ground states $|\Psi_1^-\rangle$ and $|\Psi_1^0\rangle$ with the negative (easy-axis) and positive (easy-plane) $D$, across which the fermion parity is essentially unchanged. Accompanied by the $E_{\text{YSR}}$ discontinuities or cusps of the excitation spectra at finite energies (Fig. 4f, Supplementary Fig. 18d and f), the Zeeman-split YSR peaks of distinguishing intensities switch to those of equal ones, as anticipated. Here the field $B$ breaks the equality of $|1, 1\rangle$ and $|1, -1\rangle$ in the spin-mixed state $|\Psi_1^-\rangle$, which can transmit unequally in intensity to the two excited states $|1/2, \pm1/2\rangle$. For the spin-unmixed state $|\Psi_1^0\rangle$ (or $|1, 0\rangle$), however, it always transmits to the $|1/2, \pm1/2\rangle$ states with equal intensities. We stress that the observed fermion-parity-preserving QPT is another prominent consequence of the quantum nature of impurity spins, which is directly visualized in an unprecedented manner. It differs sharply from the previously explored QPT between free-spin and partially screened states [12,16,32], across which the fermion parity alters and the cusps invariably occurs at the zero-excitation energy. Without loss of generality, the shift of $E_{\text{YSR}}$ ($\Delta E_{\text{YSR}}$) well obeys the anticipated law (Eq. 2) before the QPT, yet exhibiting larger $E$ in Fig. 4f and Supplementary Fig. 18d and f. Following the sign change of $D$ or the QPT, the energy spacing between the split YSR peaks scales nicely with the Zeeman splitting of the excited spin-1/2 state.

## 4. Summary

Our experiments, with unprecedented spatial and energy resolutions, establish a complete experimental picture on how the quantum high-spins and magnetic anisotropy interplay with the YSR bound states. Further experimental and theoretical analysis of the zero-bias pinning in the framework of the quantum many-body physics may enable a fully microscopic understanding of the robust zero-bias conductance peaks in Fe(I) and many potential systems made of Majorana fermions [35,36]. Moreover, our findings of a highly tunable $E_{YSR}$ on various transition-metal adatoms and adsorption sites, as well as the unprecedented fermion-parity-preserving QPT show that the $K_3C_{60}$ superconductor serves as a promising platform for controlling the YSR states with three independent knobs: magnetic field, adsorbed species and sites. In these contexts, YSR chains on $K_3C_{60}$ may hold promise for designing novel coupled quantum spins [3,4,37,38], in which the magnetic anisotropies and YSR bound states can be fully engineered at atomic scale.

**Conflict of interest**

The authors declare that they have no conflict of interest.

**Acknowledgments**

We thank D. E. Liu for helpful discussions and L. L. Wang for her kind support during the experiments. This work is financially supported by grants from the National Key Research and Development Program of China (No. 2022YFA1403100, No. 2017YFA0304600), the Natural Science Foundation of China (No. 12141403, No. 52388201), the Suzhou Science and Technology Program (SJC2021009) and Nano-X from the Suzhou Institute of Nano-Tech and Nano-Bionics (SINANO), the Chinese Academy of Sciences.

**Author contributions**

Can-Li Song, Fang-Sen Li, Xu-Cun Ma and Qi-Kun Xue conceived the project. Shu-Ze Wang, Xue-Qing Yu, Li-Xuan Wei, Li Wang, Kun Peng and Yu Liu synthesized and characterized the samples. Shu-Ze Wang, Xue-Qing Yu and Can-Li Song jointly analyzed the experimental data assisted with Qiang-Jun Cheng and Fang-Jun Cheng. Can-Li Song, Xu-Cun Ma and Qi-Kun Xue wrote the manuscript with comments from all authors.

**Appendix A. Supplementary data**

Supplementary data associated with this article can be found in the online version.

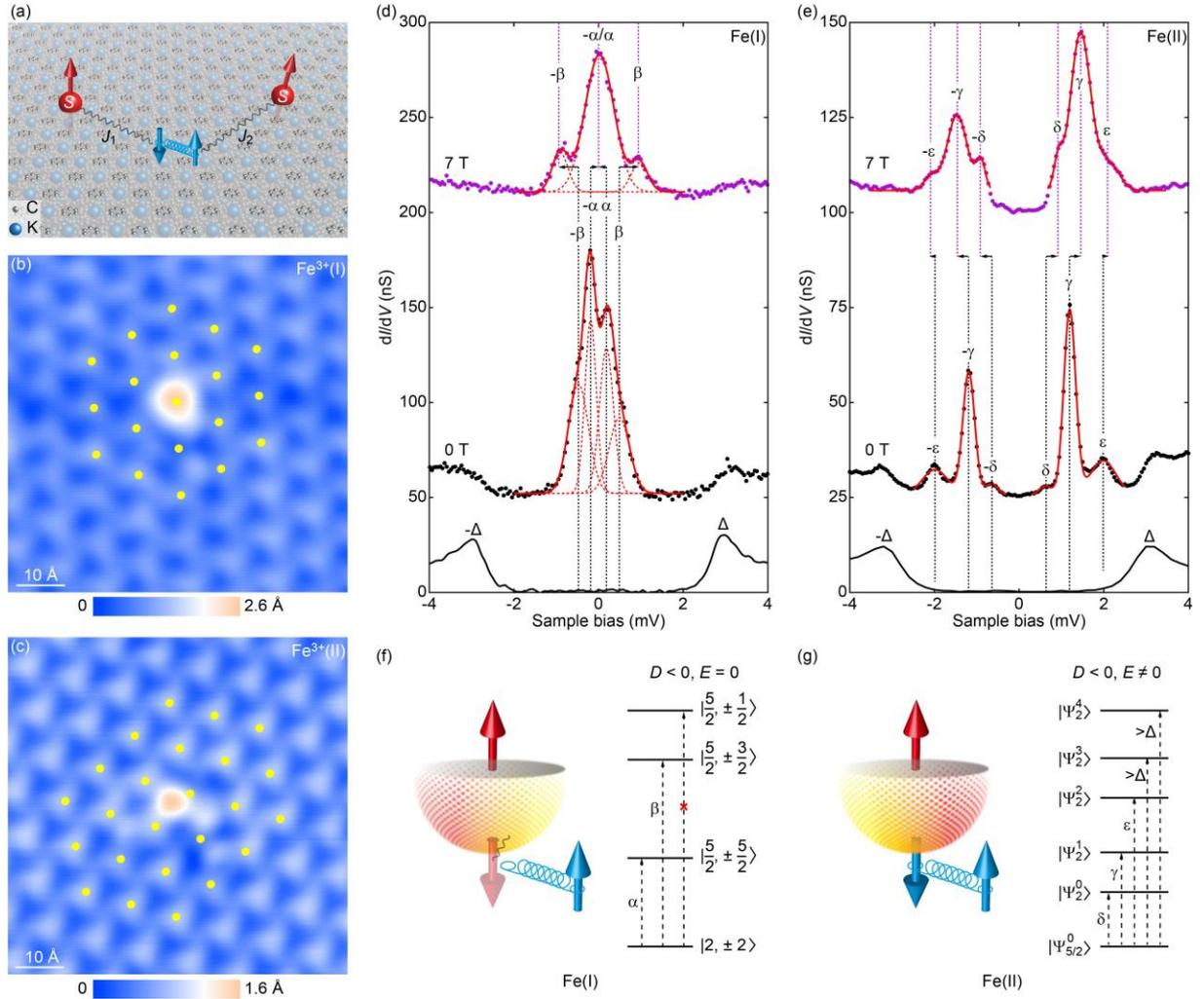

**Fig. 1. Quantum impurities on fulleride films.** (a) Schematic view of magnetic adatoms on $K_3C_{60}(111)$. (b,c) STM topographies ($V$ = 2.0 V, $I$ = 20 pA, 70 Å × 70 Å) of two distinct Fe adatoms on the top (I) and near-hollow (II) sites, respectively. Yellow dots denote the top $C_{60}$ molecules throughout. (d) d$I$/d$V$ spectra (black dots) measured at 3 Å from the Fe(I) center, showing two YSR pairs of ±α and ±β near $E_F$. The bottom spectrum marks a superconducting gap of $K_3C_{60}$ far away from the impurity atom. Red curves denote multi-Gaussian fits of the YSR peaks and the dashed ones each individual Gaussian peaks throughout. The vertical dashes mark the deduced $E_{YSR}$ before (black) and after (violet) application of an external magnetic field of 7 T. Subsequent spectra are vertically offset for clarity. Setpoint: $V$ = 15 mV and $I$ = 200 pA. (e) Same as (d) but with three YSR pairs (±δ, ±γ and ±ε) on Fe(II). (f,g) Simple scheme of magnetic anisotropy-induced splitting of Fe impurity spins and YSR multiplets. The Ψ symbols denote eigenstates of the mixed spin states.

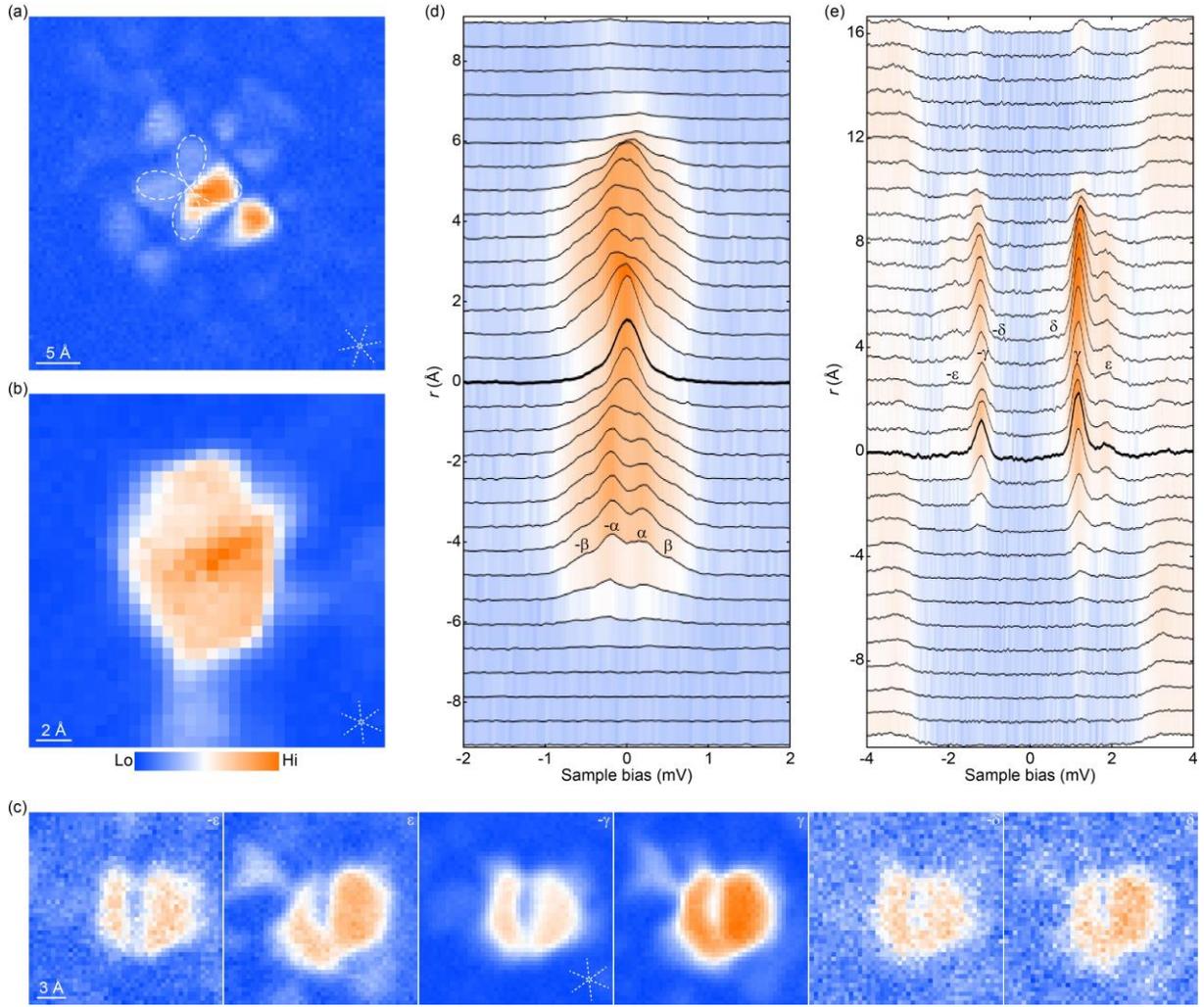

**Fig. 2. Spatial mapping of YSR wave functions under zero magnetic field.** (a) Zero-bias conductance map acquired around Fe(I) ($V$ = 30 mV, $I$ = 200 pA, 40 Å × 40 Å), imprinting with a pseudo-fourfold symmetric YSR quasiparticle cloud at 4.7 K. Overlaid is one ($d_{x^2-y^2}$ or $d_{xy}$) of the clover-leaf shaped $d$ orbitals. The three crossing lines mark close-packed directions of $C_{60}$ throughout. (b) Same as (a) but with a symmetry-reduced YSR pattern at 0.4 K ($V$ = 15 mV, $I$ = 100 pA, 20 Å × 20 Å). (c) Conductance maps ($V$ = 15 mV, $I$ = 100 pA, 20 Å × 20 Å) of the YSR states ±ε, ±γ and ±δ around Fe(II) adatom at 0.4 K, showing asymmetric dimer-like features irrespective of the energy. (d,e) d$I$/d$V$ spectra ($V$ = 15 mV, $I$ = 200 pA) taken at equal separations (0.6 Å and 0.9 Å) across Fe(I) and Fe(II), respectively. The bold curves are acquired on the impurity sites ($r$ = 0).

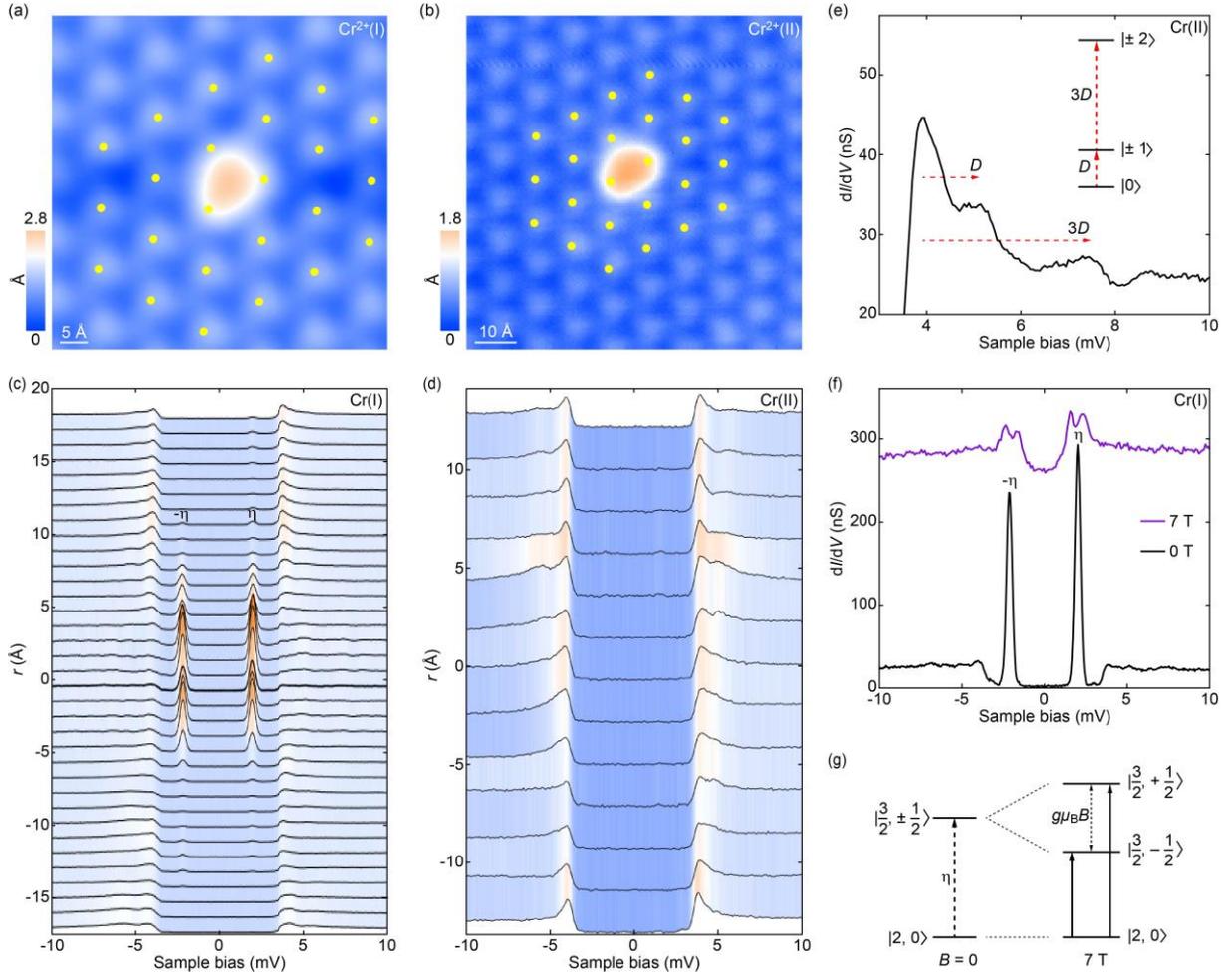

**Fig. 3. Zeeman splitting of YSR states.** (a,b) STM topographies ($V = 2.0$ V, $I = 10$ pA) of Cr adatoms bound to the bride (I, 80 Å × 80 Å) and threefold hollow sites (II, 60 Å × 60 Å) on $K_3C_{60}(111)$, respectively. (c) Zero-field d$I$/d$V$ spectra ($V = 15$ mV, $I = 200$ pA) measured at equal separations (~ 1.0 Å) across Cr(I), presenting one YSR pair (±η) near ±2.1 meV. (d) Same to (c) but at equal separations of ~ 2.0 Å across Cr(II) ($V = 15$ mV, $I = 400$ pA). Although spin-flip excitations emerge outside the superconducting gaps, no YSR state develop. (e) Zoom-in of spatially-averaged d$I$/d$V$ spectrum near Cr(II). Two conductance peaks, marked by the dashed arrows starting from the coherence peak at Δ, develop outside the superconducting gap due to inelastic spin-flip excitations for $S = 2$, as sketched in the inserted diagram. (f) d$I$/d$V$ spectra on Cr(I) in zero (black) and 7 T (violet) magnetic field. (g) Energy level diagram showing YSR excitations for the free-spin $S = 2$ state.

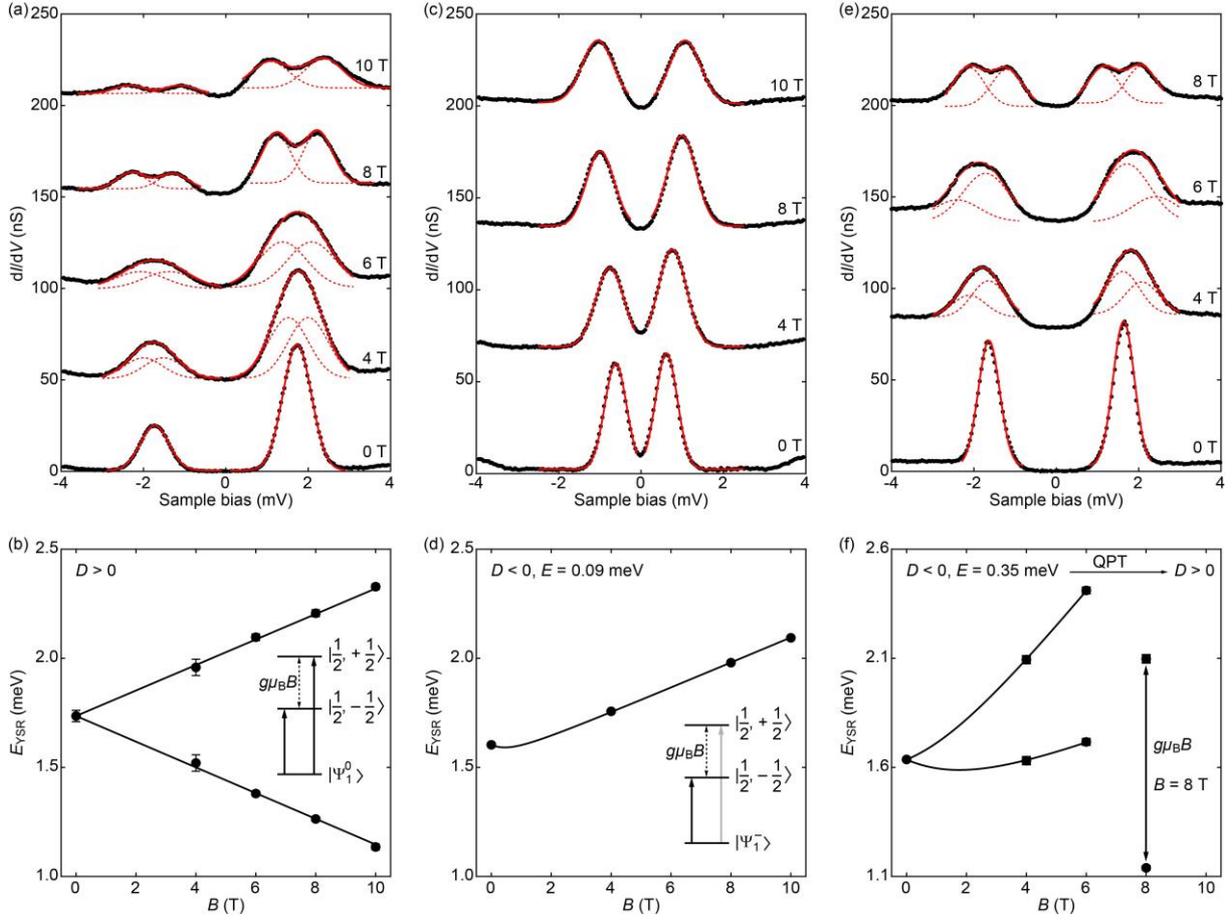

**Fig. 4. Adsorption site-dependent YSR states and QPT on Ni ($S = 1$).** (a) A series of d$I$/d$V$ spectra ($V = 8$ mV, $I = 100$ pA) under variation of an external magnetic field $B$, showing the Zeeman splitting for each YSR peak. (b) Extracted $E_{YSR}$ versus $B$, matching a free-spin ground state $|\Psi_1^0\rangle$ or positive $D$. Inserted is the energy level diagram of the corresponding YSR excitations. (c,d) Same as (a,b) but with a ground state $|\Psi_1^-\rangle$ or negative $D$ on another type of Ni impurities. In (d), the solid line shows the best fit of $\Delta E_{YSR}$ to Eq. 2, giving a transverse magnetic anisotropy $E = 0.09 \pm 0.01$ meV. (e) d$I$/d$V$ spectra ($V = 8$ mV, $I = 100$ pA) presenting a discontinuous evolution of the field-split YSR peaks in energy and intensity from $B = 6$ T to 8 T. (f) Extracted $E_{YSR}$ versus $B$, signaling a fermion-parity-preserving QPT between the ground states $|\Psi_1^-\rangle$ and $|\Psi_1^0\rangle$.

# Supplementary Information

**Section 1. Magnetic anisotropy and magnetic field effects on the YSR states**

Magnetic adatoms on superconductors often induce a pair-breaking potential for Cooper pairs and thus YSR bound states within the superconducting energy gap $\Delta$. These states carry information on the exchange coupling strength $J$ of the impurity spin $S$ with the superconductor, the magnetic anisotropies ($D$ and $E$) and external magnetic field $B$, which jointly determine the many-body ground states of the system [14, 15]. It has been established before that a competition between the Cooper pairing and the spin exchange coupling $J$ of the conduction electrons will lead to a parity-symmetry-breaking QPT between two different magnetic ground states [12, 16]. As $J$ is less than a threshold value $J_c$, the Cooper pairing overwhelms the spin screening, namely $\Delta > k_B T_K$, where $\Delta$ corresponds to the Cooper pairing strength, $k_B T_K$ ($T_K$ is the Kondo temperature) determines the energy scale of the spin screening. Consequently, the impurity is effectively decoupled from the low-energy conduction electrons and a free-spin $S$ ground state is then favored. For $J > J_c$ (or $\Delta < k_B T_K$), however, the magnetic impurity locally breaks the Cooper pairs, binding one quasiparticle to form either a fully screened Kondo singlet for $S = 1/2$ or a partially screened ground state for $S > 1/2$, i.e., $S^* = S - 1/2$.

No matter to what extent the impurity spin $S$ is screened, the system has a multiplet ground state with spin $S(S^*)$ as $S \geq 1$, from which a quasiparticle transmits to the first excited state $S^*(S)$, thereby inducing the YSR states. Without loss of generality, we choose $S_z$ as the quantization axis and the spin multiplets can be therefore labeled by its $S_z$ components. The degenerate spin multiplets will be then lifted by the uniaxial ($D$) and transverse ($E$) magnetic anisotropies and the magnetic field $B$, with the basic rules below:

1. The uniaxial magnetic anisotropy $D$ breaks the unique SU(2) symmetry and splits the spin multiplet into various sectors of $S_z = \pm S(S^*)$, $\pm(S(S^*) - 1)$, …. $\pm 1/2$ (0) for half-integer (integer) $S(S^*)$. Only $S_z$ is kept as the conserved quantity. For $S(S^*) = 1/2$, the term $D$ merely gives a trivial shift of the energy of the spin doublets.

2. The $B_z$ term further splits all spin doublets of $\pm 1/2$, $\pm 1$, …. $\pm S(S^*)$ into various $S_z$ sectors. $S_z$ remains to be the conserved quantity.

3. The transverse magnetic anisotropy $E$ mixes the component $S_z$ with $S_z \pm 2$ and the $S_z$ quantity is no longer conserved except for $S(S^*) = 1/2$.

4. The $B_x$ and $B_y$ terms mix the component $S_z$ with $S_z \pm 1$ and the $S_z$ quantity is no longer conserved for all $S(S^*) \geq 1/2$.

We illustrate the effects of $D$ and $B_z$ on the YSR bound states by the schematic plots in Supplementary Figs. 4 and 5, where the $S_z$ is the conserved quantity. For easy comparison, we mark the allowed YSR excitations ($\Delta S_z = \pm 1/2$) at zero temperature before and after application of the magnetic field $B$ (along the quantization axis $z$) by the dashed and solid arrows, respectively. Interestingly, we find that the YSR states rely crucially on the free (screened)-spin $S(S^*)$ ground state, half-integer (integer) spin and the sign of $D$, as detailed below:

1. For *D* < 0, only one pair of YSR states develops for the free-spin *S* ground state, which merely shifts away from $E_F$ by $g\mu_B B/2 = \mu_B B$ in the magnetic field *B*, irrespective of the half-integer (integer) spin *S*. For the ground states with partially screened-spin $S^*$, however, there exist two pairs of YSR states at *B* = 0. On application of the magnetic field *B*, the inner pair of YSR peaks shift towards $E_F$ by $\mu_B B$, while the outer ones shift away from $E_F$ by $\mu_B B$.

2. For *D* > 0, there are significant distinctions of the YSR states between the half-integer and integer spins. As the ground state has a free spin, there exists only one pair of YSR states that splits into two with an energy spacing of $2\mu_B B$ by the Zeeman effect of the excited spin state $S^* = 1/2$ for integer *S*, while the magnetic field *B* shifts the inner (outer) pair of YSR peaks away from (towards) $E_F$ by $\mu_B B$ for half-integer *S*. However, the opposite holds true for the screened-spin $S^*$ ground state.

These properties are highly valid for the magnetic adsorbates at high-symmetry sites (e.g., Fe(I) and Cr(I)) with a negligible *E*. Under this circumstance, there are at most two pairs of YSR peaks even with the magnetic field $B_z$ applied. More pairs of YSR peaks emerge as the transverse magnetic anisotropy *E* cannot be ignored. It will substantially mix the component $S_z$ with $S_z \pm 2$ and the spin states are therefore the mixed states except for $S(S^*) = 1/2$. As a result, the following properties are satisfied by the ground states with spin $S(S^*)$:

3. For integer $S(S^*)$, the first excited state has half-integer spin $S^*(S)$. The Kramers' degeneracy ensures that the numbers of the YSR pairs at *B* = 0 T are 1, 2, 3 for the first excited states $S^*(S) = 1/2$, 3/2 and 5/2, respectively. The external magnetic field *B* lifts the Kramers' degeneracy and consequently each YSR pair split into two. Besides, the YSR peaks will exhibit extra shifts by the $S_z$-dependent Zeeman splitting for *D* < 0. For $S^*(S) = 1/2$, the energy spacing between the split YSR pairs is $2\mu_B B$.

4. For half-integer $S(S^*)$, the first excited state has integer spin $S^*(S)$ and its degeneracy is wholly lifted in general. As thus, the numbers of the YSR pairs are $2S^*(S) +1$, irrespective of the applied *B*, if they do not merge into the continuum of quasiparticle states beyond Δ. In this situation, the magnetic field *B* merely shifts the YSR peaks, but the energy shifting will depend on *E*, rather than a simple scaling with $\mu_B B$.

5. For half-integer *S* and *D* < 0, the ground state favors a singlet $|\Psi_S^0\rangle$ with predominant weights at the largest $|S_z|$ values and the YSR pairs are all anticipated to shift away from $E_F$ by the applied magnetic field *B*. Otherwise, some YSR pairs will shift towards $E_F$.

In experiment, we invariably applied the external magnetic field *B* perpendicular to the sample surface and observed the unique Zeeman splitting of $2\mu_B B$ for $S_z^*(S_z) = \pm 1/2$. This hints that the $B_x$ and $B_y$ terms are negligibly small and the explored impurity spins tend to orient along the *B* direction. For simplicity, we constrain the discussions in the single scattering channel case, which matches nicely with our experimental observations of the identical spatial wave function for various YSR states.

To illustrate the effect of the transverse magnetic anisotropy and the external magnetic field *B* on the YSR states, we diagonalize the *S* = 1 spin Hamiltonian with $|1, -1\rangle$, $|1, 0\rangle$ and $|1, 1\rangle$ spin states, shown in Eq. 1 in the main text. Here the *E* term mixes the $|1, \pm1\rangle$ spin states to form two mixed states $|\Psi_1^\pm\rangle$, leaving $|\Psi_1^0\rangle$ = $|1, 0\rangle$ intact. Provided that the applied magnetic field *B* aligns with the spin direction ($B_z = B$, $B_x = B_y = 0$), the eigenstates and the corresponding energies can be written as follows:

$$|\Psi_1^+\rangle = \cos\gamma|1, 1\rangle + \sin\gamma|1, -1\rangle, |E_1^+\rangle = D + \sqrt{E^2+(g\mu_B B)^2}$$

$$|\Psi_1^-\rangle = \sin\gamma|1, 1\rangle - \cos\gamma|1, -1\rangle, |E_1^-\rangle = D - \sqrt{E^2+(g\mu_B B)^2}$$

$$|\Psi_1^0\rangle = |0\rangle, |E_1^0\rangle = 0,$$

where $\tan(2\gamma) = E/\mu_B B$. Apparently, the ground state with the lowest energy is either $|\Psi_1^0\rangle$ or $|\Psi_1^-\rangle$, which relies critically on the sign of $D$. As $D < 0$, the ground state ought to be $|\Psi_1^-\rangle$. Otherwise, $|\Psi_1^0\rangle$ lies in the ground state once if the $E$ and $B$ terms are not very huge. Intuitively, one can anticipate a phase transition from the ground state $|\Psi_1^0\rangle$ to $|\Psi_1^-\rangle$ with increasing $B$ for $D > 0$. Accordingly, the YSR excitations from such ground states to the first excited states $|1/2, \pm1/2\rangle$ render a parity-preserving QPT. The resultant two YSR pairs continuously evolve in energy at the QPT point when the eigenvalues $|E_1^0\rangle$ and $|E_1^-\rangle$ cross, and their intensities alter from equality to inequality. These properties contrast sharply with our experimental observation of the discontinuous evolution of $E_{YSR}$ for Ni impurities, across which the intensities of the two YSR pairs change from inequality to equality. This means that the observed QPT involves a change of the ground states from $|\Psi_1^-\rangle$ to $|\Psi_1^0\rangle$ with increasing $B$, which should be caused by a sign change of the $D$ from negative to positive. Prior to the QPT, the contrast of the two YSR pairs become more prominent in intensity with increasing $B$, due to the enhanced inequality of the $|1, -1\rangle$ and $|1, 1\rangle$ in the mixed state $|\Psi_1^-\rangle$.

**Section 2. YSR states on Ni and sign change in the local potential scattering $U$**

In experiment, we accidentally found one Ni impurity adatom that sits within one Abrikosov vortex in an external magnetic field of 8 T, where the superconducting energy gap is wholly killed. Interestingly, the tunneling d$I$/d$V$ spectrum displays an asymmetric peak near $E_F$, reminiscent of the Kondo resonance due to the spin-flip scattering (Supplementary Fig. 16). Indeed, the experimental data turns out to be nicely fitted with the Fano line shapes as expected for the Kondo resonance [1]:

$$\frac{dI(V)}{dV} \propto \frac{(q+\varepsilon)^2}{1+\varepsilon^2}, \qquad (3)$$

where $\varepsilon = (eV - E_K)/\Gamma$ is the normalized energy, $E_K$ and $\Gamma = k_B T_K$ are the energy and the half-width at half-maximum (HWHM) of the resonance, respectively. The asymmetry factor $q$ profoundly influences the line shape of the resonance from a Lorentzian peak ($q \to \infty$) over an asymmetric characteristic to a dip ($q \to 0$). It reflects the relative significance of the tunneling into the localized resonance and the delocalized band electrons. In Supplementary Fig. 16, the red line shows the best fit of the measured experimental data to Eq. 3, yielding the $E_K = 0.75$ meV, $q = 3.28$ and $\Gamma = k_B T_K \sim 3.18$ meV. The energy scale of the spin screening, $k_B T_K \sim 3.18$ meV, appears to be smaller than the usual pairing energy of $\Delta$ in $K_3C_{60}$ [24]. This echoes the free-spin ground state for the randomly distributed Ni adatoms on $K_3C_{60}$.

Nevertheless, the adsorption of the Ni impurities on the random sites of $K_3C_{60}(111)$ results into a great diversity in the energy $E_{YSR}$ and particle-hole asymmetry of the YSR states, as revealed in Supplementary Fig. 17. This is due to the adsorption site-varying exchange ($J$) and potential ($U$) interactions of Ni with the $K_3C_{60}$ superconductor. Interestingly, we reveal that the particle-hole asymmetry of the YSR states can be reversed among various Ni adatoms. Intuitively, this can be interpreted as a parity-symmetry-breaking QPT between the free ($S$)- and screened ($S^*$)-spin ground states. However, this is not true, since the screened ground state ($S^* =1/2$) of the Ni adatoms ($S = 1$) will principally induce two pairs of YSR bound states (Supplementary Fig. 5), which were never observed experimentally. Alternatively, a more straightforward explanation of the reversed particle-hole asymmetry is a sign change of the local potential scattering $U$ for different Ni adatoms, while all of them lie in the free-spin ground state of $S = 1$.

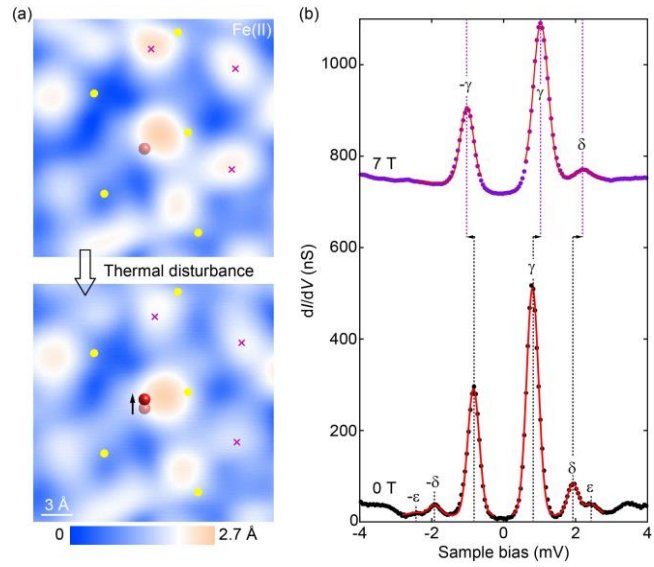

Fig. S1. (a) Zoom-in STM topographies ($V = 0.1$ V, $I = 10$ pA, 23 Å × 23 Å) of the identical Fe(II) adatom before (top) and after (bottom) a thermal circle with the sample warming from 0.4 K to 4.2 K and then cooling back to 0.4 K, showing impurity site-sensitive YSR bound states. The semi-transparent and opaque red spheres mark the Fe(II) impurity site, deduced by the simultaneously acquired YSR d$I$/d$V$ maps before and after the thermal circle, respectively. A tiny spatial movement of the Fe(II) site is visible that changes apparently the surrounding STM contrast highlighted by the fuchsia crosses. (b) Tunneling conductance d$I$/d$V$ spectra of the site-modified Fe(II) adatom under zero (black dots) and 7 T (violet dots) magnetic field. Red curves denote multi-Gaussian fits of the YSR states ($\gamma$, $\delta$ and $\varepsilon$).

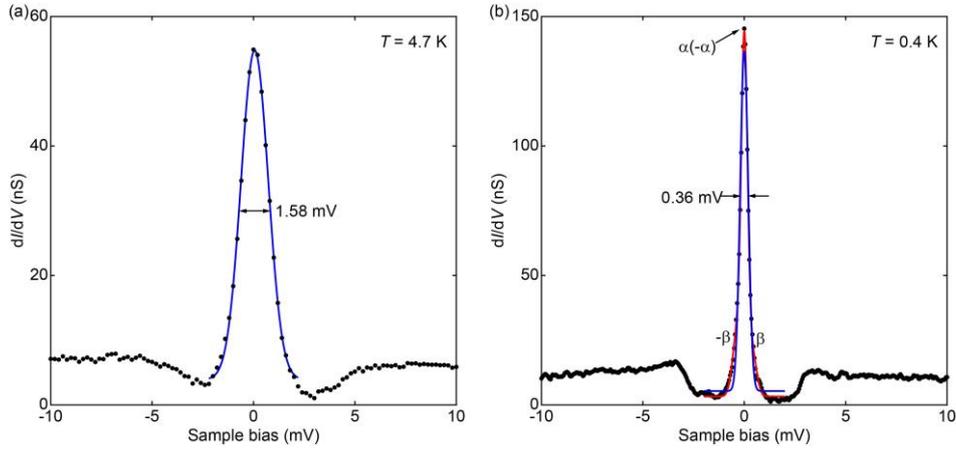

Fig. S2. (a) YSR bound states of Fe(I) impurity atom and the best fit to a single Gaussian function (blue line), showing a FWHM of ~1.58 meV at 4.7 K. (b) YSR states of the Fe(I) at 0.4 K. Blue and red lines designate the best fits to a single Gaussian function and multi (three)-Gaussian function, respectively. Evidently, the single YSR peak scenario fails to follow the experimental data (black dots) near both the bottom and top. The two pairs of YSR states ($\alpha$ and $\beta$) are spaced ~ 0.3 meV in energy and become hardly discernible by the thermal broadening at 4.7 K ($3.5k_\mathrm{B}T \sim 1.4$ meV), with $k_\mathrm{B}$ denoting the Boltzmann constant. Moreover, the energy spacing between states $\pm\alpha$ is too tiny to be resolved experimentally even at 0.4 K, giving rise to a single Gaussian peak with a FWHM of only 0.36 meV.

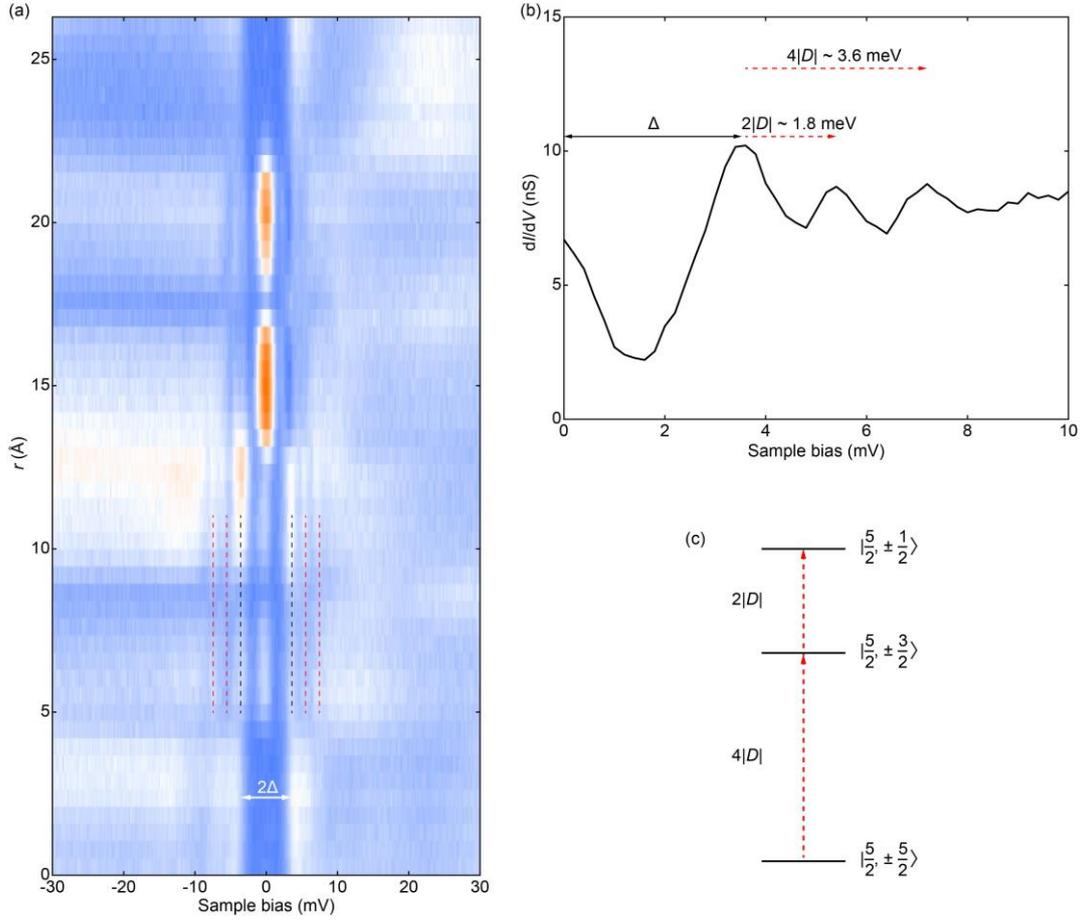

Fig. S3. (a) A series of d$I$/d$V$ spectra acquired at equal separations (0.52 Å) along one of the nodal planes of $d$ orbital-like YSR wave function of Fe(I) at 4.7 K. The YSR states exhibit an oscillatory spatial modulation in amplitude but shows no radial symmetry. Additional conductance peaks (marked by the vertical red dashes) develop outside the superconducting energy gaps (marked by the black dashes), which are derived from the inelastic spin-flip excitations. The spin-flip excitations turn out to be more pronounced on sites with attenuated YSR states. This indicates an intricate interplay between the two unconstrained events and calls for a many-body analysis of the interactions between the quantum impurity spins and the surrounding electrons. (b) Zoom-in of the spatially-averaged d$I$/d$V$ spectrum, highlighting the spin-flip structures in the empty states. Two prominent peaks develop at $\Delta + 2|D|$ and $\Delta + 4|D|$ that are consistent with inelastic spin-flip excitations for the $S = 5/2$ spin. (c) Scheme of zero-field splitting of the spin-5/2 by a uniaxial magnetic anisotropy $D < 0$. The inelastic excitation of $4|D| \sim 3.6$ meV denotes a transition from the $S_z = \pm 5/2$ to the $S_z = \pm 3/2$ state (thus $D = -0.9$ meV), while the spin-flip excitation of $2|D| \sim 1.8$ meV from the $S_z = \pm 3/2$ to the $S_z = \pm 1/2$ state is most probably caused by the spin pumping, because the thermal energy $k_B T \sim 0.4$ meV is considerably smaller than $4|D| = 3.6$ meV and it is not expected to induce observable thermal occupation of the $S_z = \pm 3/2$ intermediate state at 4.7 K. This claim matches with our observation that the inelastic spin-flip excitations become more prominent at a larger tunneling current.

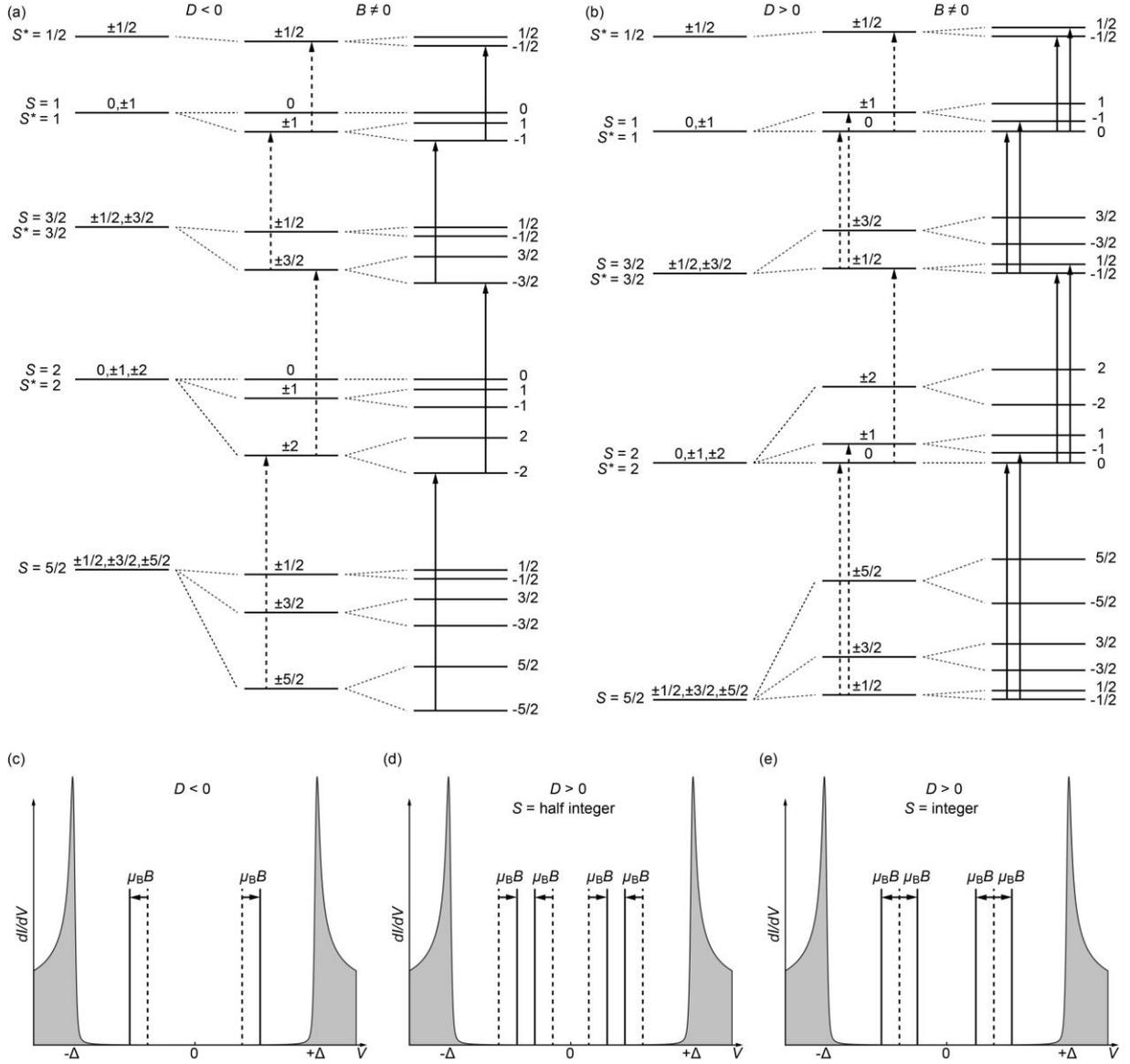

Fig. S4. YSR states of a free spin-$S$ multiplet. (a) Schematic diagram illustrating variations of the energy levels of the spin states under the easy-axis anisotropy $D < 0$ and magnetic field $B$. $S^*$ represents a partially screened impurity spin by a bound quasiparticle which acts as the excited state. The dashed (solid) arrows denote the dominant excitations ($\Delta S_z = \pm 1/2$) at zero temperature without (with) the external magnetic field. (b) Same as (a) but with the easy-plane magnetic anisotropy $D > 0$. Conversely, there exist two pairs of YSR states, even at $B = 0$ for half-integer $S$. (c) Schematic of the characteristic d$I$/d$V$ spectrum as recorded by the STM on the free spin-$S$ impurity for $D < 0$ without (dashed lines) and with (solid lines) the magnetic field $B$. In the magnetic field, we expect that the single pair of YSR peaks only shift away from $E_F$ due to the $S(S^*)_z$-dependent Zeeman splitting $\mu_B B S_z$. $\Delta$ denotes the superconducting energy gap beyond which the continuum of quasiparticle states starts. (d, e) Same as (c) but with $D > 0$. In the magnetic field $B$, the inner (outer) pair of YSR peaks shift away from (towards) $E_F$ for half-integer $S$, while each YSR peak into two for integer $S$.

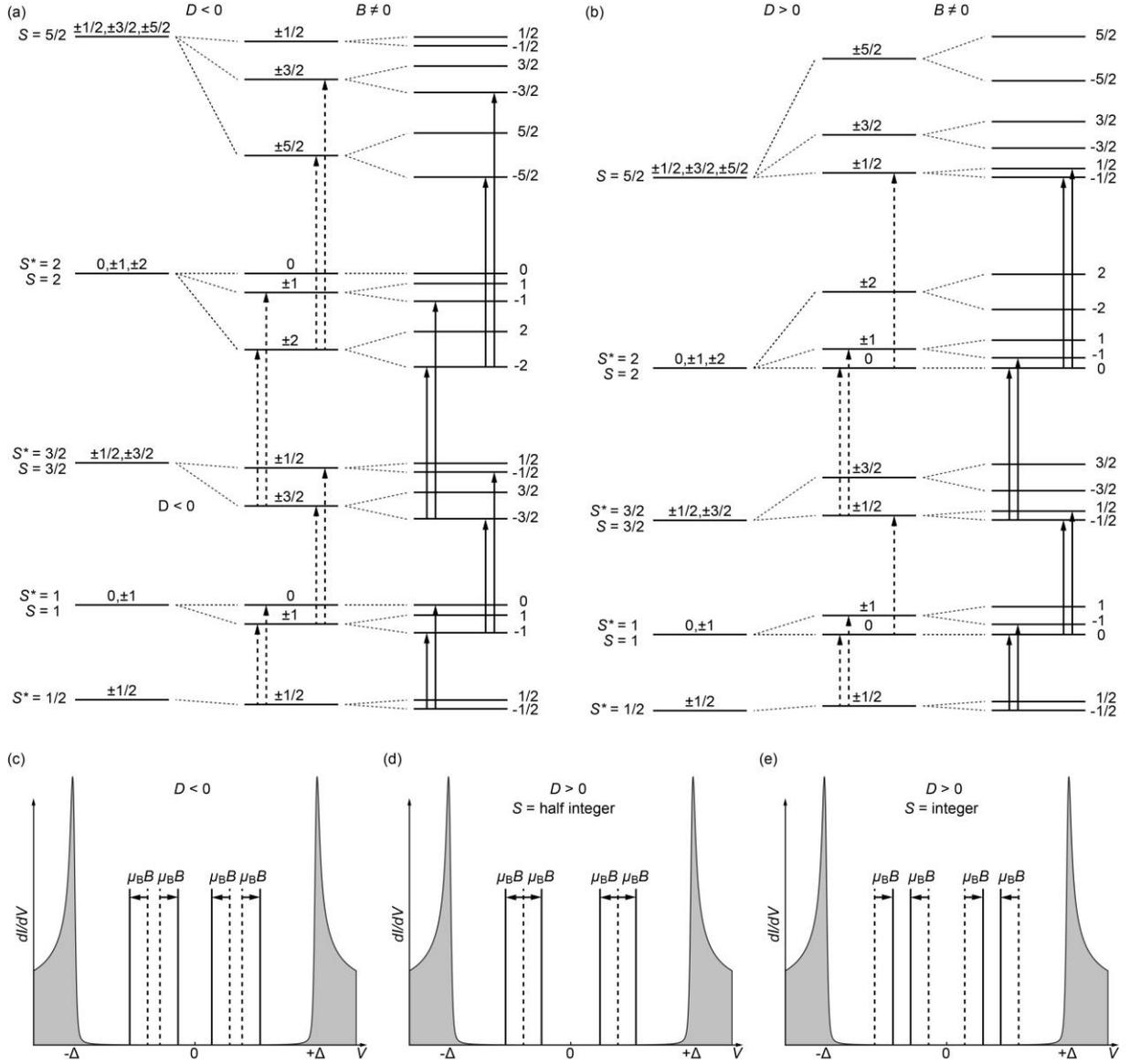

Fig. S5. YSR states of a partially screened spin-$S^*$ multiplet. (a) Schematic diagram illustrating variations of the energy levels of the spin states under the easy-axis anisotropy $D < 0$ and magnetic field $B$. Here $S^*$ represents the partially screened impurity spin by a bound quasiparticle which acts as the ground state. The dashed (solid) arrows denote the dominant excitations ($\Delta S_z = \pm 1/2$) at zero temperature without (with) the magnetic field $B$. (b) Same as (a) but with the easy-plane magnetic anisotropy $D > 0$. There exist two pairs of YSR states, even at $B = 0$ for integer $S$. (c) Schematic of the characteristic $dI/dV$ spectrum as recorded by the STM on the partially screened spin-$S^*$ impurity for $D < 0$ without (dashed lines) and with (solid lines) the magnetic field $B$. In the magnetic field, the inner pair of YSR peaks shift towards $E_F$, while the outer ones shift away from $E_F$. (d, e) Same as (c) but with $D > 0$. The magnetic field $B$ shifts the inner (outer) pair of YSR peaks away from (towards) $E_F$ for integer $S$, while it splits each YSR peak splits into two for half-integer $S$.

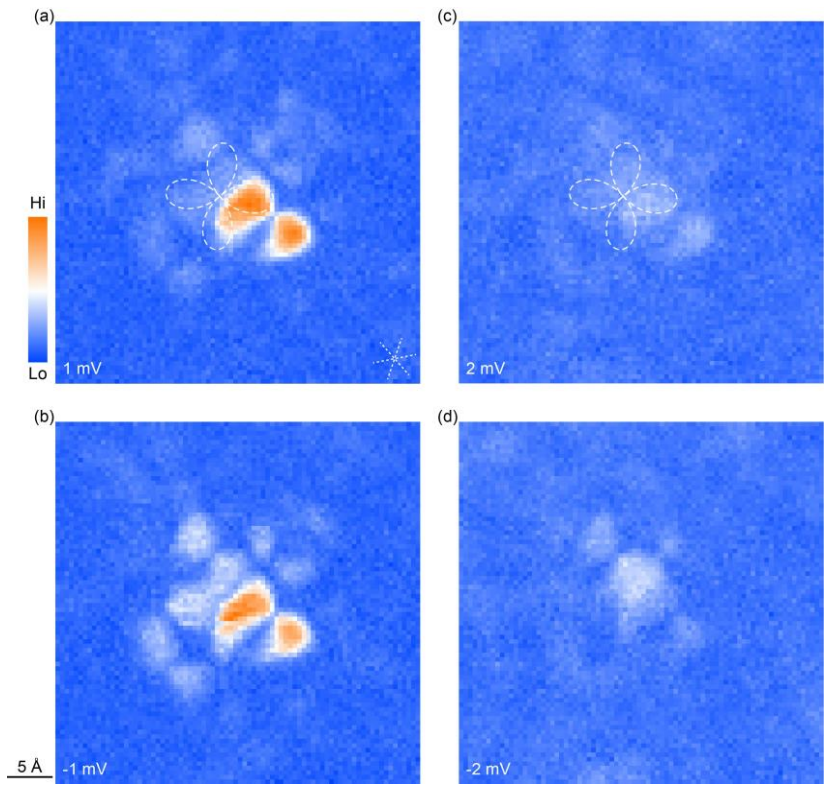

Fig. S6. (a-d) Energy-dependent d$I$/d$V$ maps in the vicinity of one Fe(I) impurity atom at 4.7 K ($V$ = 30 mV, $I$ = 200 pA, 40 Å × 40 Å), showing a fingerprint of the $d$-orbital symmetry of the YSR wave function. The quasi-four-fold symmetric YSR quasiparticle cloud is independent of the bias voltages as indicated but falls off at ± 2 mV.

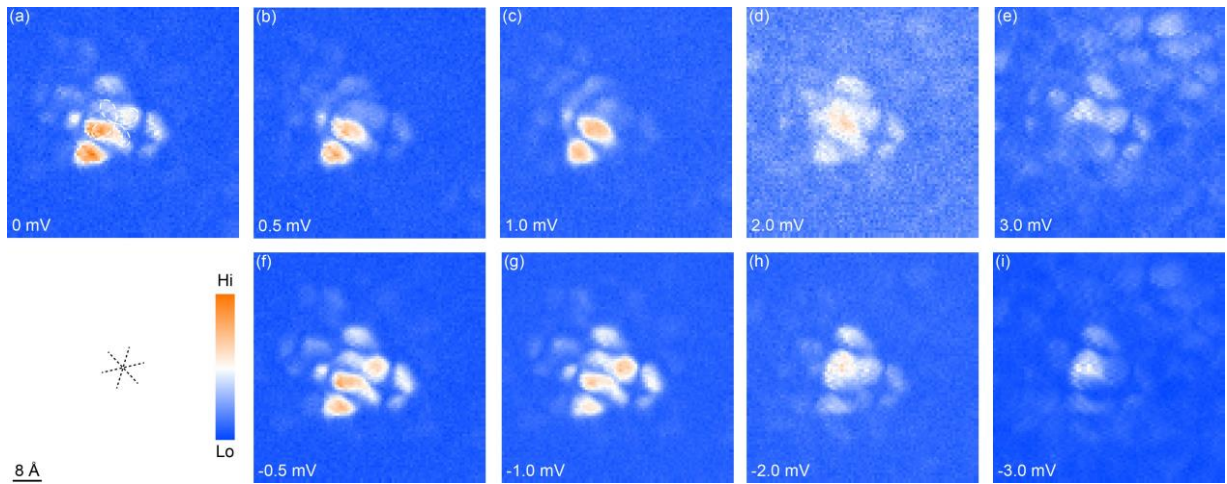

Fig. S7. (a-i) Energy-dependent d$I$/d$V$ maps in the vicinity of the Fe(II) impurity atom at 4.7 K ($V$ = 30 mV, $I$ = 100 pA, 45 Å × 45 Å), showing distorted $d$ orbital symmetry of the YSR wave function. The distortion is caused by a mismatching between the four-fold $d$ orbital symmetry and the underlying $C_{3v}$ symmetry of the K$_3$C$_{60}$(111) surface. Anyhow, one of the nodal planes of $d$ orbital-like YSR wave functions is aligned along one of the close-packed directions of the top C$_{60}$ molecules.

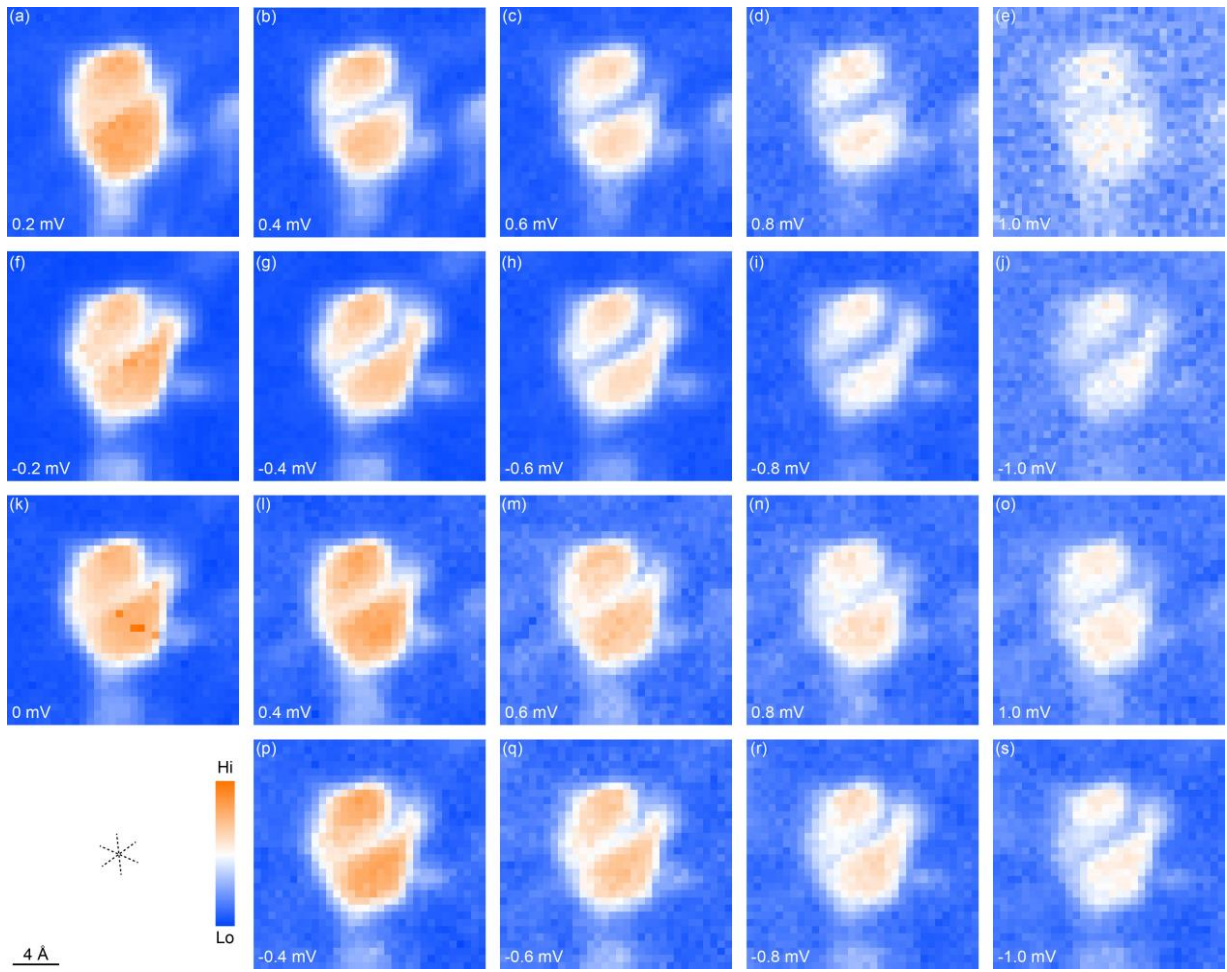

Fig. S8. (a-j) Energy-dependent *dI/dV* maps ($V$ = 15 mV, $I$ = 100 pA, 20 Å × 20 Å) in the vicinity of the Fe(I) impurity atom at $B$ = 0 and 0.4 K. (k-s) Same as (a-j) but with $B$ = 7 T. All the conductance *dI/dV* maps are characteristic of a dimer-like structure, irrespective of the sample bias voltage and polarity as indicated.

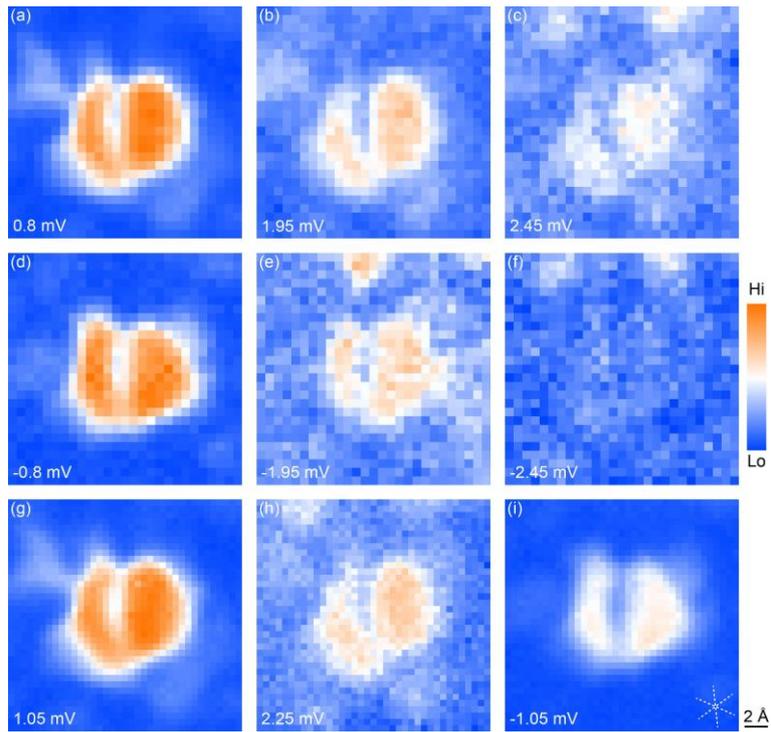

Fig. S9. (a-f) Energy-dependent d$I$/d$V$ maps ($V$ = 20 mV, $I$ = 100 pA, 20 Å × 20 Å) in the vicinity of the thermal-modified Fe(II) adatom at 0.4 K and $B$ = 0. (g-i) Same as (a-f) but with $B$ = 7 T ($V$ = 20 mV, $I$ = 200 pA, 20 Å × 20 Å). Though the $E_{\text{YSR}}$ shifts under the magnetic field $B$, the spatial distributions of their wave functions all look very alike, irrespective of the sample bias voltage and polarity as indicated. More apparently, the YSR maps exhibit no redial symmetry in space.

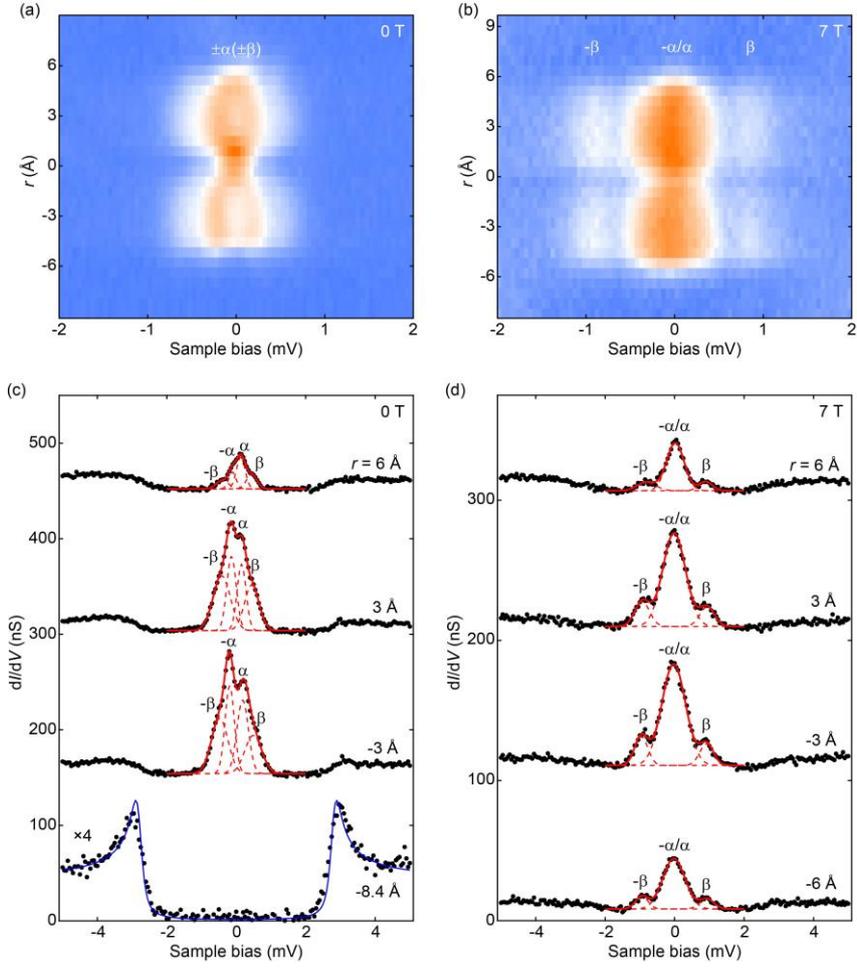

Fig. S10. (a) Color-coded and distance $r$-dependent d$I$/d$V$ spectra measured across the Fe(I) adatom at $B = 0$ and 0.4 K. (b) Same as (a) but with $B = 7$ T. The YSR states exhibit oscillatory spatial modulations in both energy and intensity. (c, d) Tunneling d$I$/d$V$ spectra (black dots) measured at representative distances from the impurity site ($r = 0$) without and with a 7 T magnetic field, respectively. Subsequent spectra are vertically offset for clarity. Red curves denote multi-Gaussian fits of the YSR peaks and the dashed ones each individual Gaussian peaks. Away from the impurity site, the zero-field spectra can be clearly divided into two pairs of Gaussian peaks symmetric to $E_F$ ($\pm\alpha$ and $\pm\beta$), whereas the inner pair of the YSR states ($\pm\alpha$) recoalesce into one single Gaussian peaks at $B = 7$ T. At $r = -8.4$ Å, the superconducting gap is evident under zero field and could be nicely fitted to a single isotropic $s$-wave pair function, marked by the blue curve in (c). Setpoint: $V = 15$ mV, $I = 200$ pA.

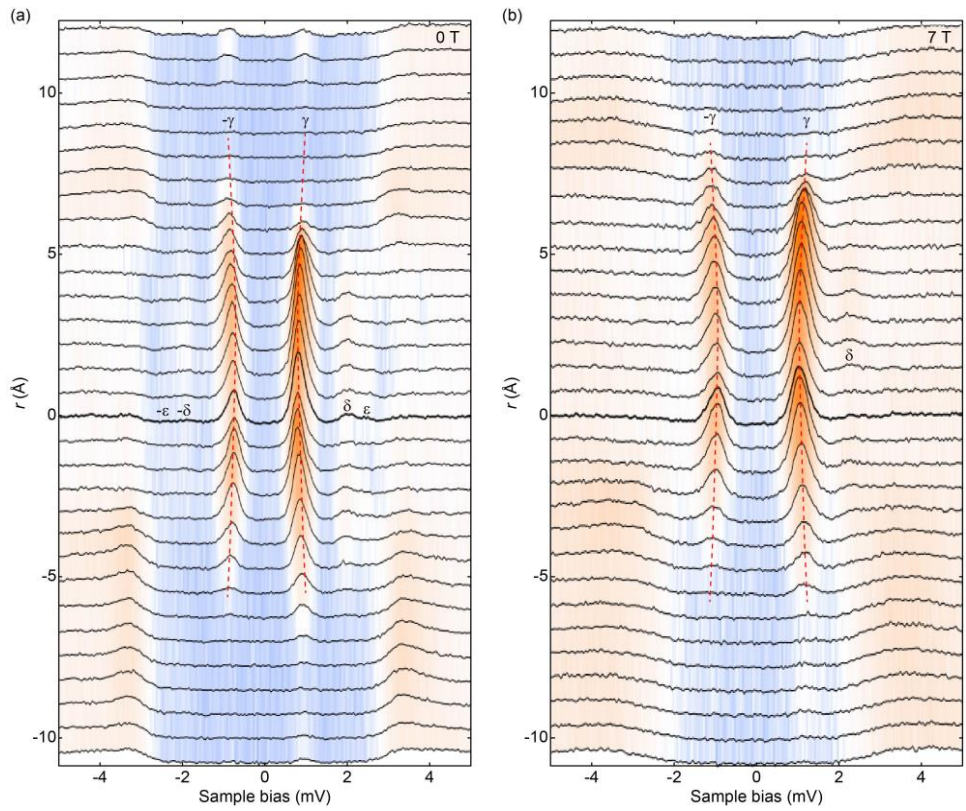

Fig. S11. (a) Color-coded and distance $r$-dependent d$I$/d$V$ spectra measured across the thermal-modified Fe(II) adatom at $B = 0$ and 0.4 K. (b) Same as (a) but with $B = 7$ T. The red dashed lines track the spatial evolution of the YSR states $\pm\gamma$. At 7 T, the $\pm\varepsilon$ and -$\delta$ YSR states become too faint to be resolved in experiment. Setpoint: $V = 15$ mV, $I = 200$ pA.

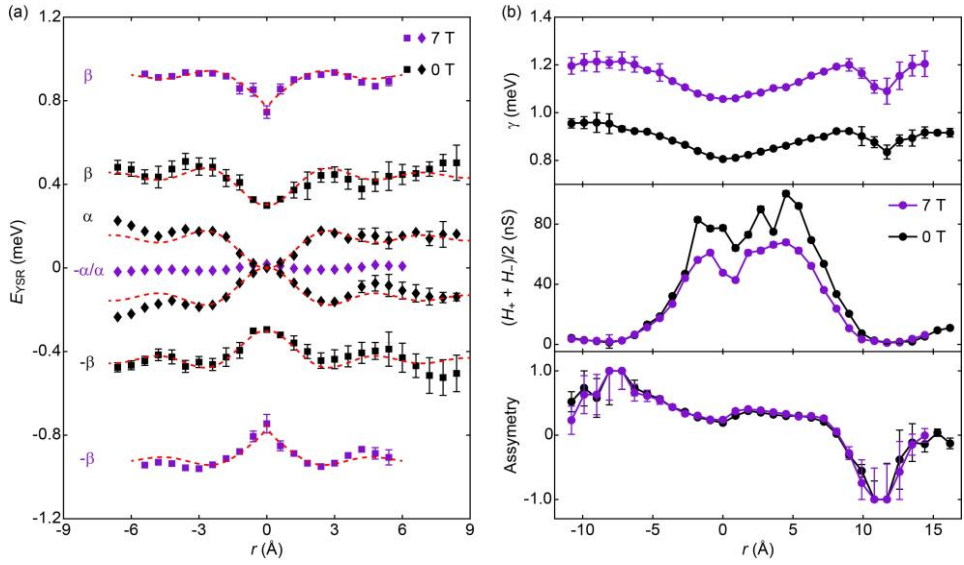

Fig. S12. (a) Dependence of the YSR energy $E_{YSR}$ on distance $r$ from the Fe(I) impurity site ($r = 0$) under zero (black symbols) and 7 T (violet symbols) magnetic field, measured along the nodal plane of the $3d$ orbital-like YSR wave function. The red dashes approximately track the spatial evolution of the YSR energy. (b) Energy (top), mean peak height $(H_+ + H_-)/2$ (middle) and asymmetry $(H_+ - H_-)/(H_+ + H_-)$ (bottom) of the YSR resonance $\pm\gamma$ plotted as a function of distance $r$ from the Fe(II) impurity site ($r = 0$) under zero and 7 T magnetic field, taken along the anti-nodal plane of the $d$-orbital YSR wave function. Peak heights of the holelike ($H_+$) and electronlike ($H_-$) components of the YSR bound states are extracted from Gaussian fits to the YSR resonance peaks $\pm\gamma$ above and below $E_F$. Unlike the $E_F$-symmetric energy, the YSR heights $H_\pm$ are distinct at positive and negative sample biases, and their asymmetry $(H_+ - H_-)/(H_+ + H_-)$ displays a striking spatial dependence that breaks the radial symmetry of the YSR states. Away from the impurity site, the asymmetry increases to unity and then decreases at $r < 0$, whereas it overall declines and changes sign at $r > 0$. The error bars correspond to the standard deviations from the Gaussian fits and are mostly smaller than the symbol size.

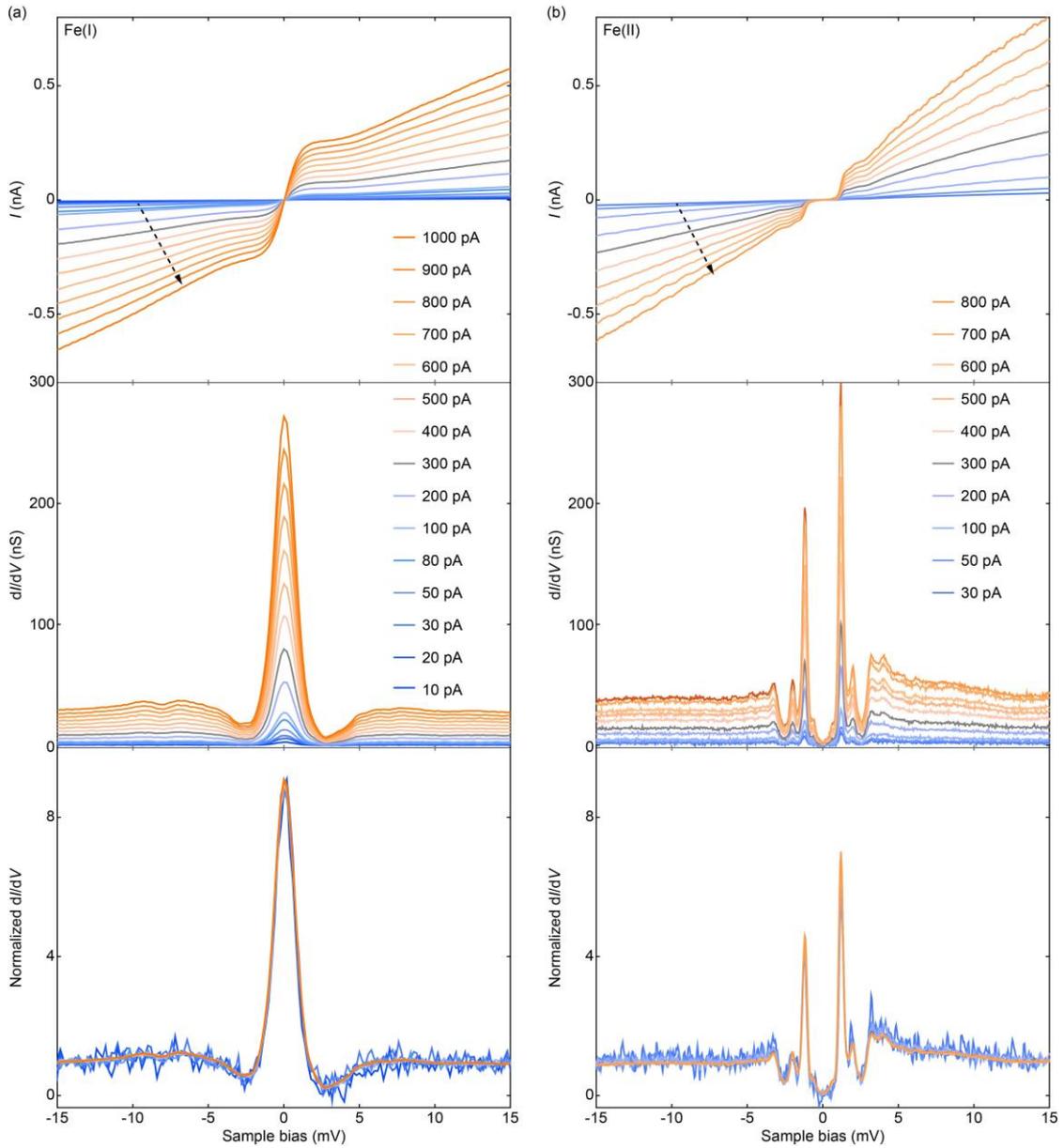

Fig. S13. (a) Current $I$ (top), raw (middle) and normalized (bottom) differential conductance d$I$/d$V$ spectra on Fe(I), measured as a function of increasing tip-to-sample distance. The current setpoint is changed from $I = 10$ pA (large distance, light blue curve) to 1000 pA (short distance, orange line) at a constant sample voltage $V = 30$ mV, as marked by the black arrow. (b) Same as (a) but on Fe(II), with the current setpoint changing from $I = 30$ pA (large distance, light blue curve) to 800 pA (short distance, orange line) at a constant sample voltage $V = 15$ mV. The normalization was performed by dividing the raw d$I$/d$V$ spectra by their respective conductance values at $V = 30$ meV in (a) and 15 meV in (b), respectively.

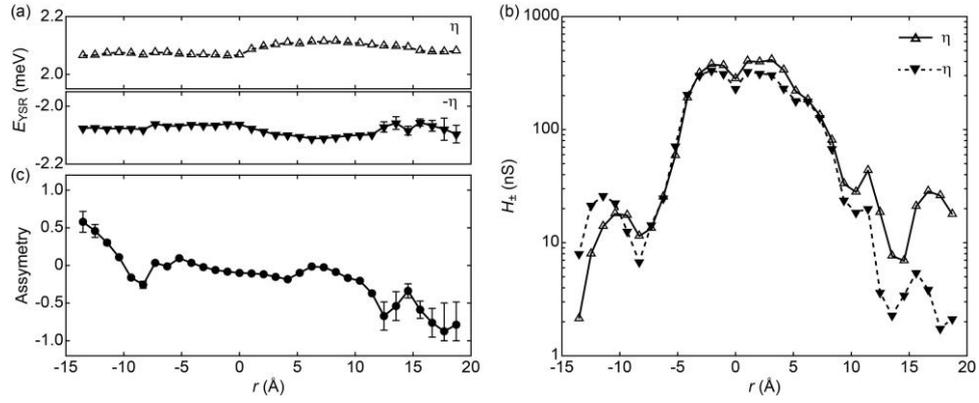

Fig. S14. (a) Dependence of the YSR energies of $\pm\eta$ on distance $r$ from the Cr(I) impurity site ($r = 0$). (b) YSR peak height $H_+$ (empty up-pointing triangles, $\eta$) and $H_-$ (solid down-pointing triangles, $-\eta$) as a function of the distance $r$ from the Cr centre. (c) Particle-hole asymmetry of the YSR states $\pm\eta$. Peak heights $H_\pm$ are extracted from Gaussian fits to the YSR peaks $\pm\eta$.

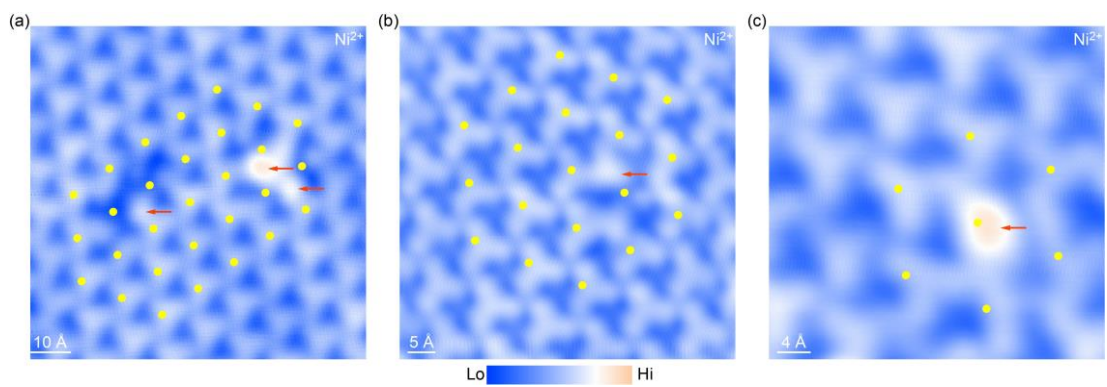

Fig. S15. (a-c) Representative STM topographies ($V = 1.5$ V, $I = 30$ pA) showing random distribution of Ni adatoms on $K_3C_{60}$, marked by the red arrows. In contrast with the Fe and Cr adatoms, the Ni adatoms are more randomly adsorbed on the $K_3C_{60}$(111) surface. Image size: (a) 80 Å × 80 Å; (b) 60 Å × 60 Å; (c) 40 Å × 40 Å.

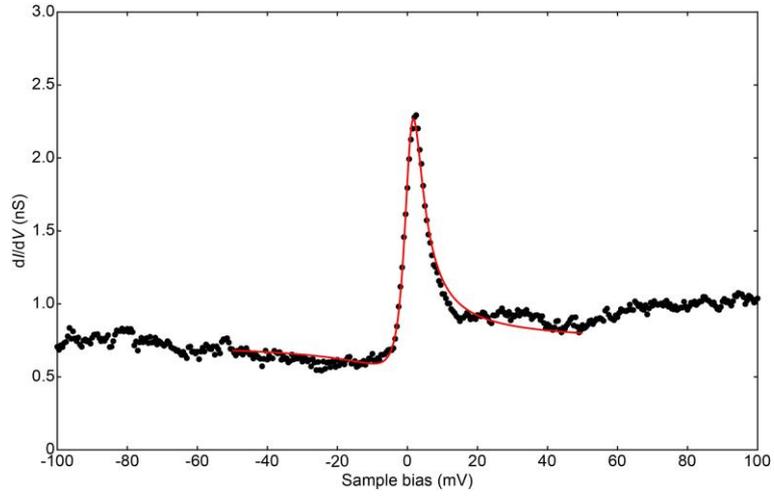

Fig. S16. Tunneling d$I$/d$V$ spectrum (black dots) measured on one Ni impurity adatom that happens to sit within one Abrikosov vortex in a magnetic field of 8 T, displaying an apparent Kondo resonance. The red line designates the best fit of the data to the Fano model in Eq. 3, giving rise to a Kondo screening energy scale $k_\text{B}T_\text{K}$ ~ 3.18 meV < $\Delta$. Set point: $V$ = 100 mV, $I$ = 100 pA.

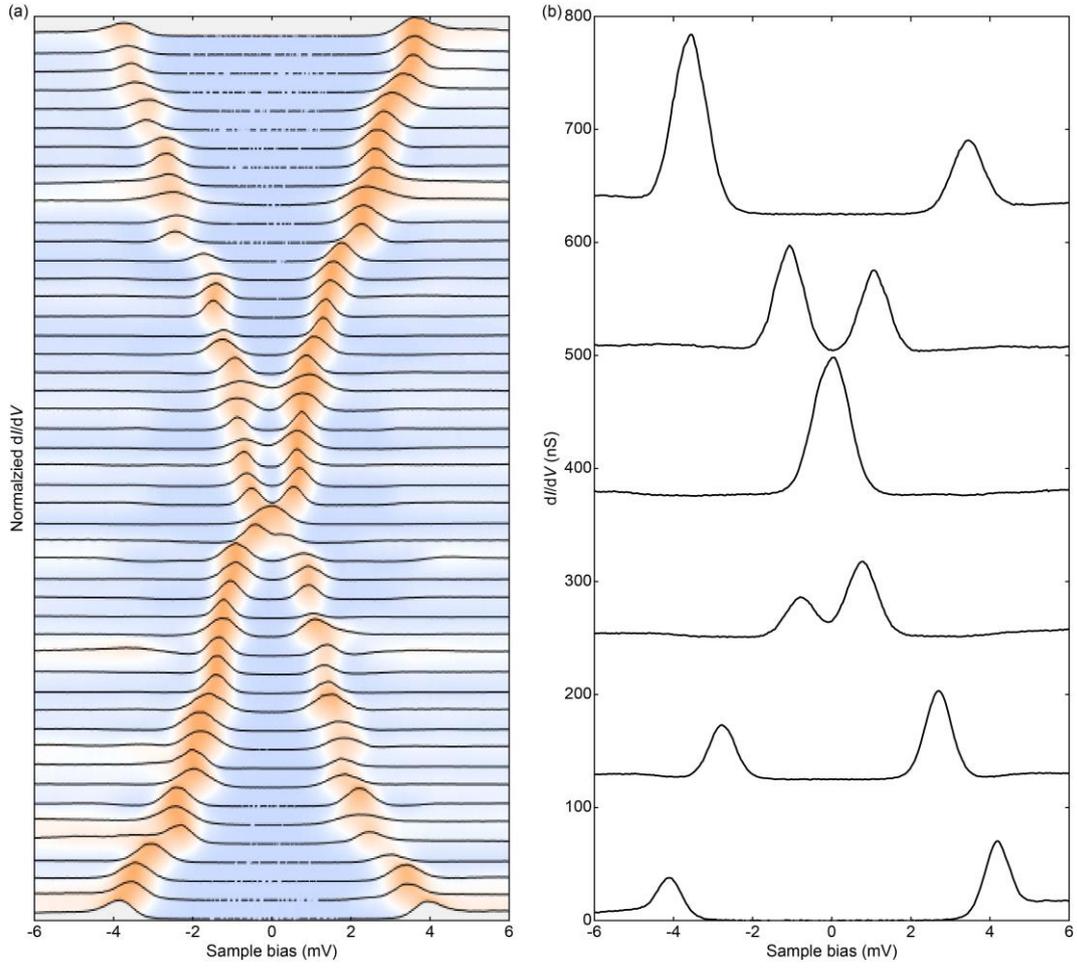

Fig. S17. (a) Color-coded plot of d$I$/d$V$ spectra on 48 representative Ni impurity adatoms, all showing one pair of YSR states at 0 T. For comparison, we normalize every d$I$/d$V$ spectrum by dividing it by its higher YSR peak and sort them by the energy of the stronger YSR peak. (b) Raw d$I$/d$V$ spectra with apparent particle-hole asymmetry. Note that the particle-hole asymmetry relies crucially on the registry of the Ni adatoms and can be reversed, due to the local variations and sign change in the potential scattering $U$. Setpoint: $V = 8$ mV and $I = 100$ pA.

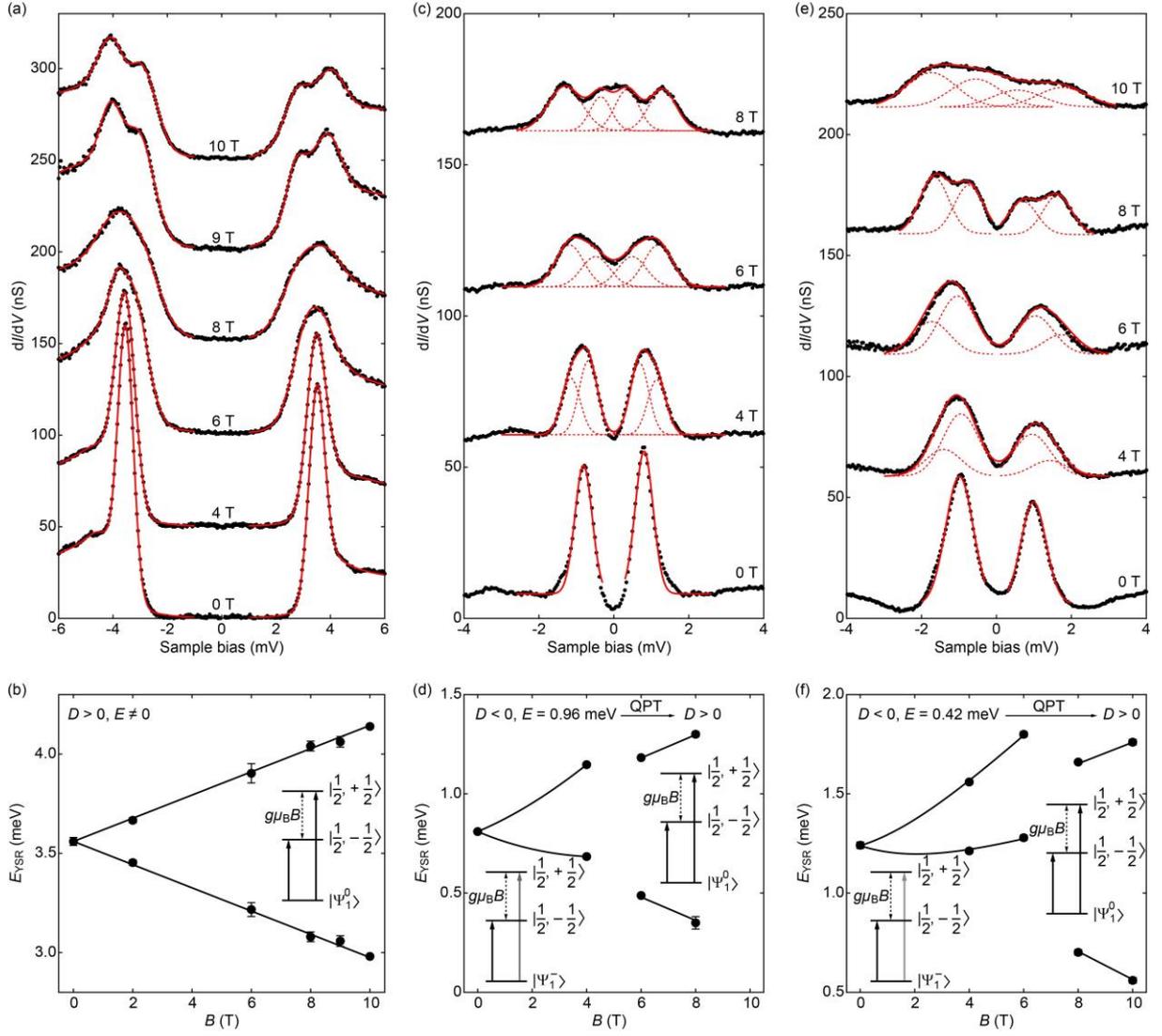

Fig. S18. (a) Additional d$I$/d$V$ spectra ($V$ = 8 mV, $I$ = 200 pA) plotted as a function of the magnetic field $B$, showing the Zeeman splitting for each YSR peak on Ni ($S$ = 1). Throughout this figure, the YSR peaks are stronger in the occupied states. This contrasts with the YSR states in Fig. 3, which we attribute to a sign change in $U$. (b) Extracted $E_{YSR}$ versus $B$, consistent with a free-spin ground state $|\Psi_1^0\rangle$ or positive $D$, irrespective of the sign change in $U$. (c-f) Same as (a, b) but with a ground state $|\Psi_1^-\rangle$ or negative $D$ on another two Ni impurity adatoms ($V$ = 8 mV, $I$ = 100 pA). No matter how the particle-hole asymmetry or the $U$ changes, the fermion-parity-preserving QPT from the $|\Psi_1^-\rangle$ to $|\Psi_1^0\rangle$ ground states can be frequently observed as the magnetic field $B$ is increased. Accordingly, the Zeeman-split YSR peaks of distinguishing intensities switch to those of equal ones. In (d, f), the solid lines show the best fit of $\Delta E_{YSR}$ to Eq. 2, yielding the transverse magnetic anisotropy $E$ = 0.96 meV and 0.42 meV, respectively. Inserted are the energy level diagram of the corresponding YSR excitations.

22